\documentclass{emulateapj}
\usepackage{natbib}

\newcommand{\degree}{\ensuremath{^\circ}}
\newcommand{\kmskpc}{\ensuremath{{\rm km\:} {\rm s}^{-1} {\rm kpc}^{-1}}}
\newcommand{\meanv}{\ensuremath{<v>}}
\begin{document}

\title{Radial Dependence of the Pattern Speed of M51}
\author{Sharon E. Meidt and Richard J. Rand}
\affil{Department of Physics and Astronomy, \\University of New Mexico, 800 Yale Blvd Northeast, Albuquerque, NM 87131}
\author{Michael R. Merrifield}
\affil{School of Physics $\&$ Astronomy, \\University of Nottingham, University Park, Nottingham, NG7 2RD, UK}
\author{Rahul Shetty}
\affil{Center for Astrophysics\\Harvard University, Cambridge, MA 02138}
\author{Stuart N. Vogel}
\affil{Department of Astronomy, \\University of Maryland, College Park, MD 20742-2421}

%
%

\begin{abstract}
The grand-design spiral galaxy M51 has long been a crucial target for theories of spiral structure.  Studies of this iconic spiral can address the question of whether strong spiral structure is transient (e.g. interaction-driven) or long-lasting.  As a clue to the origin of the structure in M51, we investigate evidence for radial variation in the spiral pattern speed using the radial Tremaine-Weinberg (TWR) method.  We implement the method on CO observations tracing the ISM-dominant molecular component.  Results from the method's numerical implementation--combined with regularization, which smooths intrinsically noisy solutions--indicate two distinct patterns speeds inside 4 kpc at our derived major axis PA=170$\degree$, both ending at corotation and both significantly higher than the conventionally adopted global value.  Inspection of the rotation curve suggests that the pattern speed interior to 2 kpc lacks an ILR, consistent with the leading structure seen in HST near-IR observations.  We also find tentative evidence for a lower pattern speed between 4 and 5.3 kpc measured by extending the regularized zone.  As with the original TW method, uncertainty in major axis position angle (PA) is the largest source of error in the calculation; in this study, where $\delta_{PA}$=$\pm$5$\degree$, a $\sim$20\% error is introduced to the parameters of the speeds at PA=170$\degree$.  Accessory to this standard uncertainty, solutions with PA=175$\degree$ (also admitted by the data) exhibit only one pattern speed inside 4 kpc, and we consider this circumstance under the semblance of a radially varying PA.  
\end{abstract}

\keywords{galaxies: spiral --
galaxies: kinematics and dynamics -- 
galaxies: structure --
methods: numerical}
\section{Introduction}
\indent  The large angular size and clear spiral structure of the nearly face-on spiral M51 make it ideal for studies of the nature and origin of grand design spiral structure. Two scenarios dominate the discussion in the literature, each based on opposing theories: strong spiral structure as a quasi-stationary density wave (e.g. \citealt{linshu}), or as a transient feature due to interaction with nearby companion NGC 5195 (e.g. \citealt{tully}).  \\
\indent Observations of both the stellar and gaseous components reveal consistencies with the density wave interpretation at some level (see \citealt{elmegreen1}; \citealt{vog93}; \citealt{rand93}).  In accord with the seminal study of \citet{tully} which attributes the (transient) outer pattern to the interaction with its companion and proposes that the inner arms are likely spiral density waves also driven by the encounter, \citet{elmegreen1} and \citet{vog93} find independent evidence for two different pattern speeds.  If the strong spiral's corotation radius overlaps with the ILR of the outer, material pattern, as is suspected (e.g \citealt{tully}; \citealt{elmegreen1}), this may indicate the stimulation of an inner spiral wavemode by the outer material spiral via mode-coupling \citep{tagger}.  But observationally it remains unclear whether, if driven by the outer pattern, the strong spiral structure is transient, or has survived a few rotation times (e.g. as speculated by \citealt{vog93}).  \\
\indent In the simulations of both \citet{salo} and \citet{howardByrd} structure throughout the disk is well reproduced by multiple passages of the (bound) orbiting companion, and the nuclear structure, in particular, seems intimately related to the inward propagation of multiple tidally-induced perturbations \citep{salob}.  Pursuant to the study of \citet{toomresq}, such simulations of M51 have proved indispensable for exploring and motivating scenarios in favor of short-lived waves.  In the short-lived wave paradigm, propagating wave packets evolved from kinematic distortions in the outer disk may be swing-amplified \citep{toomre}, causing a strong response in the inner disk.  As predicted by \citet{salob}, the wave speed has a complex radial dependence, featuring a constant pattern speed for the dominant $m$=2 structure out to 1.2-1.8 kpc (depending on the disk mass assumed), followed by a superposition of structures described by a pattern speed that decreases with radius, down to $\sim$10-20 $\kmskpc$ by $\sim$4.6 kpc.  \\
\indent Knowledge of the pattern speed of the structure can, in principle, both distinguish between and reconcile the short- and long-lived wave scenarios, and so many studies have focused on measuring and characterizing this parameter (see \citealt{elmegreen1}, \citealt{tully}, \citealt{salob} and \citealt{garcia}, for example).  The pattern speed of the outer spiral has long been proposed near 10-20 $\kmskpc$.  In the inner disk, application of the traditional, model-independent method of Tremaine \& Weinberg (1984; hereafter TW) using CO observations yields a pattern speed $\Omega_p$=38 $\kmskpc$ \citep{zrm04}, in general agreement with the determinations based on resonance locations of \citet{elmegreen1} and \citet{tully} (but higher than the pattern speed $\Omega_p$=27 $\kmskpc$ found by \citealt{garcia}).  Although the TW analysis shows evidence for significant departures from the expected relation for a constant pattern speed in both the inner- and outer-most regions of the disk, the method cannot quantitatively account for any suspected radial variation of the pattern speed.\\
\indent The radial TW (hereafter TWR) method (\citealt{mrm}; \citealt{meidt}) should prove an invaluable resource in this regard, since with it we can characterize the angular speeds of distinct patterns and their possible radial variation.  For the first time, we are able to observationally address issues related to the complex nature and persistence of spiral patterns and the connection, if any, between multiple pattern speeds in a single disk.  \\
\indent Like its traditional counterpart, the TWR method, summarized in $\S$ 2, relies on the use of a kinematic tracer found to obey continuity. Here, we consider the ISM-dominant molecular component in the inner disk of M51 as traced by CO observations.  In $\S\S$ \ref{sec:obs} and \ref{sec:molecdomin} we describe these observations and review the arguments which establish their conformity with the assumptions of the method.  In $\S$ \ref{sec:quad} we formulate the TWR quadrature and motivate the models developed for testing.\\
\indent Results of the regularized TWR calculation applied to the inner disk are presented in $\S$ \ref{sec:inneronly}.  There we establish a best estimate for the pattern speed(s) of the bright spiral structure by considering the characteristics of solutions over a $\pm$5$\degree$ range of disk position angles (PAs) ($\S$ \ref{sec:PAchar}); according to the findings of \citet{debPA}, we can expect uncertainty in the PA to be the dominant source of systematic error in the calculation.  We also compare this estimate to other tested models of the radial dependence in $\S$ \ref{sec:state}.  \\
\indent In an effort to authenticate the TWR estimate, in $\S$ \ref{sec:corrollary} we relate our measurements to other independent evidence for more than a single pattern speed in the inner disk of M51 and investigate the resonance locations and overlaps that they entail.  We also consider our measurements in light of relevant findings throughout the literature, including those of \citet{shetty} ($\S$ \ref{sec:varyPA}) and \citet{henry} ($\S$ \ref{sec:discussion}).  Final results are summarized in a conclusion section.
\section{\label{sec:TWRreg}The TWR method with Regularization}
The radial modification of the TW method (\citealt{mrm}; \citealt{meidt}, hereafter M08) delivers a derivation for radially-dependent pattern speeds measurable from observationally accessible quantities.  The so-called TWR calculation proceeds under assumptions parallel to those of the original method, namely that the disk of the galaxy is flat (unwarped); that the surface density of a disk component, which must obey continuity, becomes negligibly small at some radius and all azimuths within the map boundary (thereby critically yielding converged integrals; see below); and that the relation between the emission from this component and its surface density is linear, or if not, suspected deviations from linearity can be modeled. \\
\indent Departure from the traditional method (which assumes that the disk contains a single, well-defined rigidly rotating pattern) emerges by allowing that $\Omega_p$=$\Omega_p(r)$ -- and the surface density of the tracer $\Sigma (x,y,t)=\Sigma(r,\phi-\Omega_p(r)t)$. Integration of the continuity equation obeyed by the tracer thereupon yields a Volterra integral equation of the first kind for $\Omega_p(r)$,
\begin{equation}
\int_{r=y}^\infty \left\{[\Sigma(x',y)-\Sigma(-x',y)]r\right\}\Omega_p(r)dr=\int_{-\infty}^\infty \Sigma v_y dx
\label{eq:volt}
\end{equation}
where $x'(r,y)=\sqrt{r^2-y^2}$ \citep{mrm}.  This equation can be cast in terms of $x_{obs}$ and $y_{obs}$, the coordinates in the plane of the sky along the major and minor axes, respectively, and $v_{obs}$, the observed l.o.s. velocity, since for a galaxy projected onto the sky plane with inclination $\alpha$, $x=x_{obs}$, $y=y_{obs}/\cos \alpha$, and $v_y=v_{obs}/\sin \alpha$.\\
\indent When the integral on the left of equation (\ref{eq:volt}) is replaced with a discrete quadrature for different values of $y=y_i$ and $r=r_j$ (represented in Figure 1 of M08), equation (\ref{eq:volt}) takes the form of the matrix expression
\begin{equation}
 K_{ij}\Omega_j=b_i 
\label{eq:twr}
\end{equation}
with $\textbf{K}$ an upper triangular $N\times N$ square matrix.  This can be solved numerically for a total of two independent measures of $\Omega_p(r)$, one from either side of the galaxy (y$>$0 and y$<$0).\\
\indent As described by \cite{mrm} (and depicted in Figure 2 of M08), solving Equation \ref{eq:twr} by standard back-substitution results in the propagation of errors from large radii inward, whereby solutions inescapably display noisy oscillations.  As demonstrated there, applied first to Sb galaxy NGC 1068, this effect can be impeded most simply by adopting a relatively large bin width; the TWR solution in this case is found to decrease with radius, and yield a winding time estimate for the two-armed structure.  \\
\indent But in general, as found in application to simulations (M08), smaller radial bins are preferable to insure accurate assessment of radial variation.  In addition, noisy behavior in solutions tends to be amplified when the quadrature extends out to the edge of the surface brightness (a requirement argued for by M08), which not least imposes that the outermost bins generally cover the lowest S/N regions in the disk.  Combined with a relatively small bin, numerical solutions as a result of inward error propagation display a systematic offset in each bin between measurement and the actual value, preventing accurate determination of $\Omega_p(r)$ (M08).\\
\indent As shown in M08, regularization provides an effective means of reducing the intrinsic propagation of noise in solutions while maintaining the precision required to accurately identify true radial variation.  There, regularized TWR calculations were applied successfully to simulated disks featuring multiple pattern speeds in distinct radial zones as well as spiral winding.  \\
\indent Following M08, then, we introduce a regularizing operator, or smoothing functional $\textbf{S}$, containing a priori information in the manner of Tikhonov-Miller regularization (\citealt{tich}; \citealt{mill}) into the $\chi^2$ estimator minimized by solutions $\Omega_j$ of equation (\ref{eq:twr}), whereby minimization returns smoothed solutions according to (in matrix form) 
\begin{equation}
(\mathbf{\bar{K}}^T\cdot\mathbf{\bar{K}} +\lambda \mathbf{S}) \cdot\mathbf{\Omega} =\mathbf{\bar{K}}^T\cdot\mathbf{\bar{b}}
\label{eq:twrreg}
\end{equation}
where the elements of $\bar{\textbf{K}}$ and $\bar{\textbf{b}}$ are $K_{ij}/\sigma_i$ and $b_i/\sigma_i$, respectively (with errors $\sigma_i$ representing the measurement error of the $i^{th}$ data point $b_i$), and the parameter $\lambda$ controls the degree of smoothness achieved in solutions.  Details for the full calculation and analysis can be found in M08; we proceed by highlighting only a few of the main precepts.\\
\indent By incorporating simple expectations from theory and observation into the smoothing $\textbf{S}$, the solution of equation \ref{eq:twrreg} yields smoothed, testable models for $\Omega_p(r)$. These models we restrict to simple forms and consider only polynomial solutions with constant, linear and quadratic radial dependence. (The elements of the smoothing $\textbf{S}$ are associated with the minimization of the $n^{th}$ derivative of $\Omega(r)$ for each polynomial solution of order $n$.)  These polynomial models can be incorporated into step-functions which parameterize the radial domains of multiple pattern speeds (see M08).  \\
\indent The best fit global solution constructed from the average of like-model solutions from the two sides is established using the standard $\chi^2_\nu$ ($\chi^2$ per degree of freedom) statistic, as in M08.  (Note that an explicit assumption here is that all patterns in the disk are indeed global.)  To summarize, once equation (\ref{eq:twrreg}) is solved with a set of prescribed smoothings on each side, we use equation (\ref{eq:twr}) to generate a complete set of $<$$v$$>_i=b_i/(\int \Sigma dx)_i$ for each global model.  The $\chi^2_\nu$ fit of the model-reproduced to actual $<$$v$$>_i$ given global measurement error $\sigma^{<v>}$ (defined as the average of the individual errors $\sigma_i^{<v>}$ for each slice) is then calculated for each.  For this $\chi^2_\nu$, we adopt the uniform weighting scheme advocated by M08.\\
\indent According to the M08 prescription where measurement errors for each slice reflect random noise in the data, for this analysis we assign errors $\sigma_i^{<v>}$ that define the change in the measured $<$$v$$>$ introduced by a change in the chosen flux cut-off in the first moment map.  Specifically, the error in the $<$$v$$>$ measured from a map with an $n\sigma$ level cutoff is defined as 
\begin{equation}
\footnotesize \sigma^{\meanv}=\frac{1}{\sqrt{2}}
\left[(\meanv_{(n-1)\sigma}-\meanv_{n\sigma})^2
+(\meanv_{(n+1)\sigma}-\meanv_{n\sigma})^2 \right]^{1/2} 
\end{equation}
for each slice $i$. The average of the individual errors then define the global measurement error across the entire disk where $\sigma^{<v>}$=$\left(\sum_{i=1}^{2N}\sigma_i^{<v>}\right)$/2N (and $N$ is the number of bins/slices used in the TWR calculation on a single side).  (Note that this error relates to the error $\sigma_i$ for each slice in equation (\ref{eq:twrreg}) through $(\int \Sigma dx)_i$.) \\
\indent Although these random errors are used in the goodness-of-fit criterion, the overall error in the measurement of $\Omega_p(r)$ given by the best-fit global model solution is defined relative to systematic errors in the calculation.  Uncertainty in the assumed position angle (PA), for example, has the largest potential for introducing errors into $<$$v$$>_i$, or conversely, the $b_i$ in equation (\ref{eq:twrreg}), and is the dominant source of error in TW calculations (\citealt{debPA}; M08).  We assess this error by testing the sensitivity of the solutions to departures from the nominal value for the PA (or inclination, for instance).  Unless otherwise specified, in this paper all reported error bars reflect the influence of PA uncertainty alone.  As for inclination errors, apart from the change introduced in the pattern speed measurements through a change in $\sin{\alpha}$, these prove to be of little additional consequence to the accurate placement of radial bins defined in the quadrature (as suggested by M08), despite the relatively low inclination (we adopt $\alpha$=24$\degree$; see Table \ref{tab-params}) of the disk of M51.  We therefore do not report this error, but instead note that a change in the inclination by $\delta_{\alpha}$=$\pm$3$\degree$ corresponds to a fractional variation of about 12$\%$ in the pattern speed estimates reported here.
\section{\label{sec:applying}Application to M51}
\subsection{\label{sec:obs}Observations}
\begin{figure}
\begin{center}
\epsscale{1.15}
\plottwo{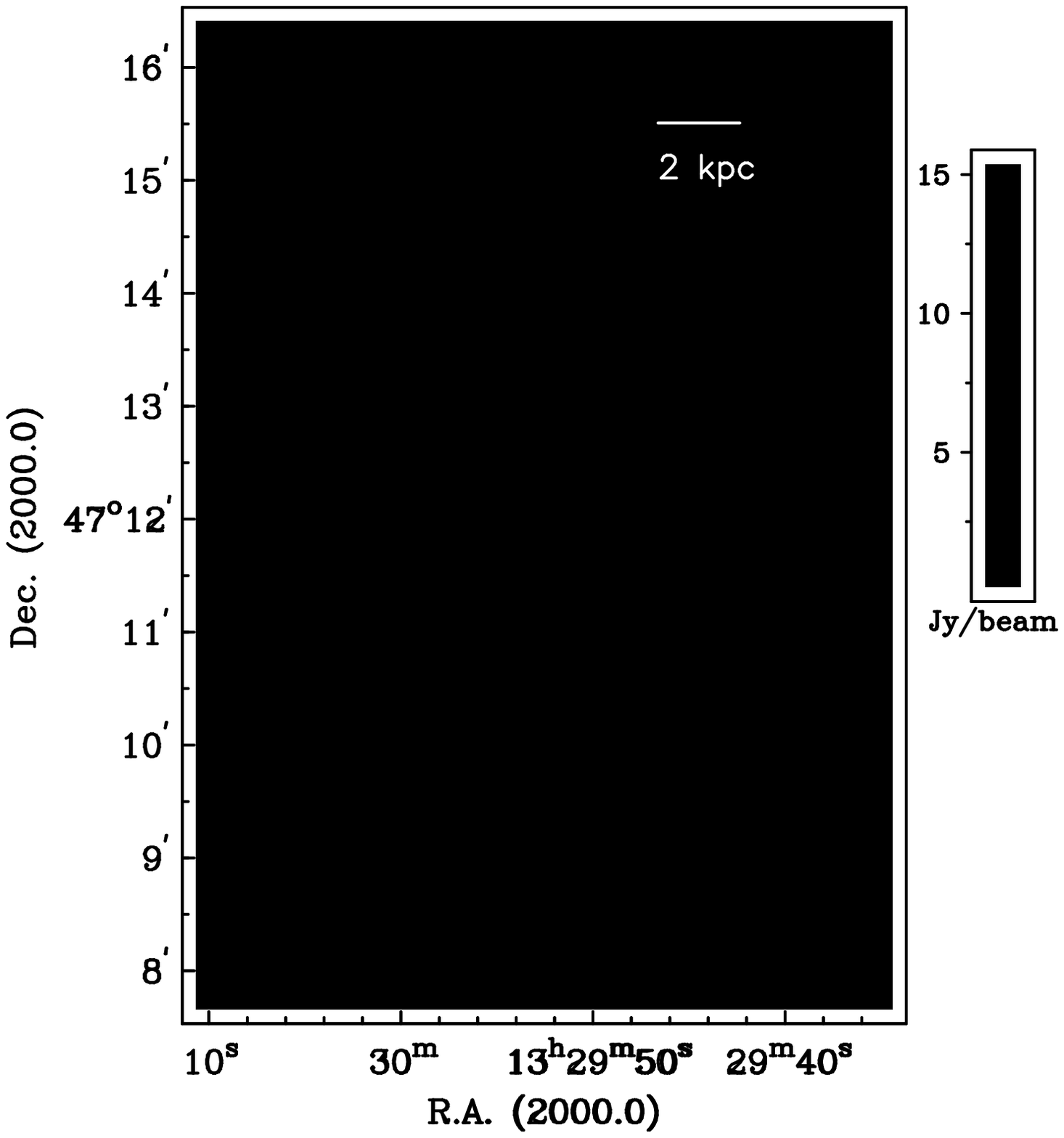}{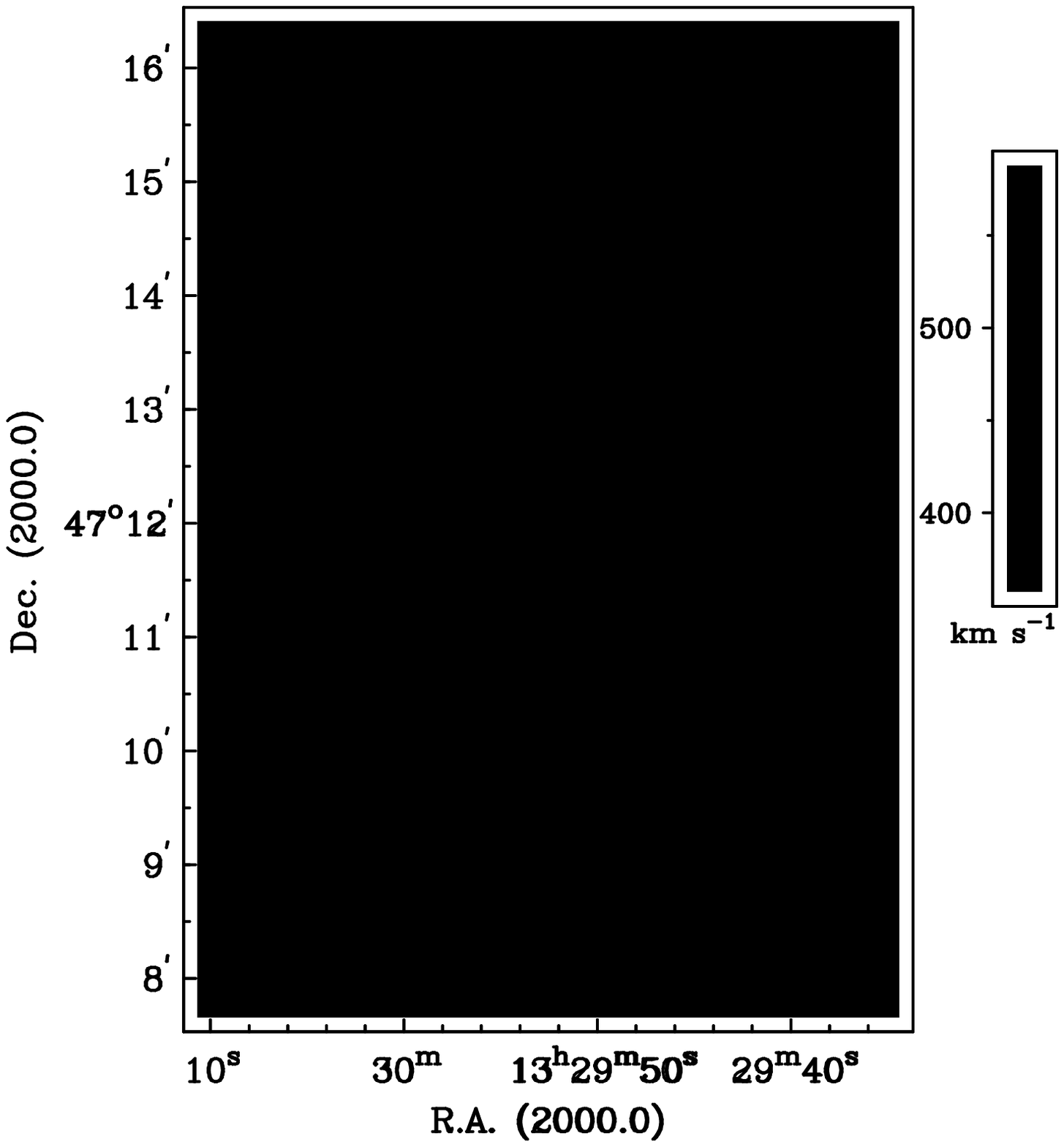}
 \end{center}
    \figcaption{Zeroth (top) and first (bottom) moment maps of the M51 CO cube (originally presented in \citealt {shetty}).  The horizontal bar in the top right corner indicates the physical scale.  The y$>$0 quadrature generally covers the eastern half of the image (depending on the value of the PA adopted), while the y$<$0 quadrature covers the western. 
\label{fig-mom0}}
\end{figure}
\indent In this paper, we consider the disk of M51 traced by high resolution CO observations.  As described in its initial publication \citep{shetty}, the cube consists of the BIMA Survey of Nearby Galaxies (SONG) observation together with several additional pointings which extend the map out to $r$$\sim$280'' and provide higher angular resolution in the central regions (see the beginning of $\S$ \ref{sec:quad}).  A complete description of the data can be found in \citet{shetty}.  The 2-$\sigma$ zeroth and first moment maps used in this analysis, derived from the full cube, are shown in Figure \ref{fig-mom0}.  Measurement errors given by uncertainty in the flux-cutoff  are defined relative to maps at the 1- and 3-$\sigma$ levels.
\subsection{\label{sec:molecdomin}Establishing Molecular Dominance}
The measured intensities and velocities in Figure \ref{fig-mom0} are suitable for use with the TWR method provided that the assumptions listed in the previous section are satisfied.  While the continuity requirement can be particularly limiting, \citet{zrm04} and \citet{rw04} argue that CO emission, the standard tracer of the molecular component of the ISM, suitably meets the TW assumptions for galaxies where the ISM is everywhere dominated by molecular gas.  This is founded on the low true efficiency of star formation in spirals, which implies that only a small fraction of molecular gas is converted into stars on orbital timescales, while molecular dominance implies that the conversion of molecular hydrogen into other phases of the ISM occurs at low levels.  \\
\indent \citet{zrm04} applied the TW method under this premise, showing with CO and HI observations that the gas content of M51 is in fact dominated by molecular hydrogen where CO is detected.  The CO observations used in this work can be similarly asserted to obey continuity: assuming a conversion factor between CO intensity and H$_2$ column density $X$=2$\times$10$^{20}$cm$^{-2}[$ K km s$^{-1}]^{-1}$, molecular hydrogen is found dominant over the majority of the CO emitting disk (roughly $R$$<$105''), where N(H$_2$)/N(H1)$\sim$10 \citep{shetty}. \\
\indent The possibility of variation in the CO-H$_2$ conversion factor has also been addressed by \citet{zrm04}, who find in a series of tests applied to M51 that neither a linear relationship between metallicity and $X$-factor nor arm-interarm variations at levels suggested by \citet{garcia} produce a significant change in the derived pattern speed estimate.  \\
\indent The negligible effect of radial dependence in $X$ can be largely attributed to the cancellation of axisymmetry with TW integration along each slice (see \citet{zrm04} for a complete account).  Analogously, for the TWR method, as long as the metallicity changes negligibly over the width of a radial bin, we can expect little change in the results of calculations that assume an approximately constant $X$-factor throughout the disk.  We have confirmed this to be the case here; modeled according to the \citet{bresolin} metallicity gradient 0.02 dex kpc$^{-1}$, an increase in $X$ with radius produces negligible change in the measured solutions.  \\
\indent An arm-interarm contrast as suggested by \citet{garcia}, on the other hand, in general may not so readily translate from TW to TWR calculations inconsequentially.  Currently, assessing the particular effect of azimuthal variation in the $X$ factor on the results of the TWR method is beyond the scope of this work.  Here, we rely on the results of \citet{zrm04} to assert that, to a first approximation, variation in $X$ should not compromise the analysis as presented here.  
\subsection{\label{sec:quad}Defining the quadrature and developing testable models}
\begin{table}
\begin{center}
\caption{Parameters used in the TWR calculation.\label{tab-params}}
\begin{tabular}{rccc}
\tableline\tableline

 Parameter& Value\\
\tableline
 Dynamical Center RA ($\alpha$) (J2000)&13$^h$29$^m$52$^s$.71\\
 Dynamical Center DEC ($\delta$) (J2000)&47$\degree$11'42''.80\\
 Distance&9.5 Mpc&\\
 Systemic Velocity (V$_{sys}$)&469 km s$^{-1}$ (LSR)\\
 Position Angle &170$\degree$$\pm$5$\degree$\\
 Inclination &24$\degree$$\pm$3$\degree$\\
\tableline
\end{tabular}
\tablecomments{The dynamical center and inclination angle are adopted from \citet{shetty}.  Entries for $V_{sys}$ and PA originate from the tilted ring analysis of the first moment of the CO cube using the GIPSY task ROTCUR.}
\end{center}
\end{table} 
\indent In order to achieve as accurate a quadrature as possible and also limit errors caused by the misdesignation of any transitions (e.g. given the finite bin width to which solutions are confined; M08), we adopt a radial bin width $\Delta r$=0.23 kpc ($D$=9.5 Mpc).  This corresponds to the limiting resolution ($\sim$4'') of the map at the innermost radii.  Since with the majority of our analysis of M51 we are most interested in characterizing the pattern speeds of the bright spiral structure, this choice is expected to yield high quality solutions for the pattern speeds in this region in particular.  \\
\indent At the largest radii (and in interarm regions) the resolution decreases to 6''-13'' \citep{shetty}. Though in principle the quadrature can accommodate a non-uniform bin width, we maintain $\Delta r$=0.23 kpc throughout the disk and rely instead on the allocation of information administered by regularization.  We assert that, even with our 4.5'' radial bins, regularized TWR calculations are prevented from oversampling the data as long as any distinct regions parameterized by the models are larger than the resolution.  \\ 
\indent As assessed in M08, we can expect departures from the nominal values of the parameters appearing in Equation \ref{eq:twr} to introduce non-negligible errors into the TWR solutions.  Uncertainty in the major axis PA is the dominant source of systematic error in the calculation, resulting in errors on the order of 20\% in TWR pattern speeds (M08).  Since the kinematic parameters of M51 are notoriously difficult to constrain, perhaps the greatest challenge to the accuracy of our solutions lies in the the accuracy with which we can constrain the quadrature.  \\
\indent To best equip the analysis in this capacity, then, we survey both our own derivations of the kinematic parameters and those from the literature.  For the coordinates of the center of rotation and the disk inclination angle, for example, we rely on the values from the study of \citet{shetty}.  These we then adopt in fits of a tilted ring model to the CO velocity field with the Groningen Image Processing System (GIPSY) program ROTCUR to determine the systemic velocity and the kinematic line of nodes (as well as the rotation velocity as a function of radius).  The resulting parameters (listed in Table \ref{tab-params}) are consistent with most previous determinations.  Note, however, that rather than adopting the range of PAs (170\degree to 180$\degree$) considered by \citet{shetty}, we initially choose PA=170$\degree$$\pm$5$\degree$.  This is principally in order that our results are more easily compared with the majority of studies on M51, especially those which entail estimates for the pattern speed.  Additionally, for this study we assume D=9.5 Mpc; with the alternative D=7.7 Mpc (more common to recent studies), all distances reported here decrease by a factor of $\sim$0.2, while all pattern speeds reciprocally increase.  \\
\indent According to the arguments in M08, we extend the unique quadrature established with the values in Table \ref{tab-params} out to the map boundary $\pm$$y_{max}$ in order to insure that all information critical for characterizing the patterns of interest is accounted for.  Since the emission extends (roughly East/West) out to $\pm$$y_{max}$=145'', this defines the maximum radial extent of the quadrature $R_{max}=y_{max}$/$\cos{\alpha}$=7.3 kpc, and hence the limit of integration along each slice $\pm$$X_i$=$\sqrt{R_{max}-y_i/\cos{\alpha}}$. Though this does place $X_{i}$ within the edge of the emission at small $\vert y\vert$ (given the elongated emission in this map from N to S and the low disk inclination), the radial range of the quadrature is still comfortably outside the radius where the integrals converge.  \\
\indent As diagnosed by M08, with this fairly extensive quadrature solutions $\Omega_p(r)$ are at risk of regularization-induced bias.  This bias is defined for the particular case when bins cover a region that displays only faint emission, has little information from a strong pattern, or is suspected of sustaining multiple patterns; when these bins are prescribed an unrealistic model, the accuracy of the remainder of the solution can be jeopardized.  \\
\indent Evidence for regions in the outer disk of M51 susceptible to regularization-induced bias are identifiable a priori in the intensity map and its Fourier power spectrum.  Later in $\S$ \ref{sec:inneronly} where we address this bias and its signatures, we adopt the counter-measure developed by M08 wherein the compromised bins are calculated without regularization.  This imposes the additional parameterization of a cut radius $r_c$ on our model solutions, interior to which regularization proceeds as defined in equation \ref{eq:twrreg}.  (See M08 for a description of the calculation and analysis in this case.)  In practice, we unregularize only as long as we can insure the sustained effectiveness of regularization in the rest of the calculation, given that an increased number of unregularized bins promotes the reintroduction of unamendable propagating noise.  \\
\indent In order to test for the possibility of multiple pattern speeds and/or winding, models additionally parameterize either single or multiple distinct radial domains over which the solution can vary as zeroth, first or second order polynomials.  Though in general we test all possible models at each stage of the analysis, in some cases we restrict our consideration to only those polynomials for which the degree of freedom plus 3-4 bins does not exceed the number of bins in a given domain.  
\section{\label{sec:results}M51: Results}
\begin{figure}
\begin{center}
\epsscale{1.0}
\plotone{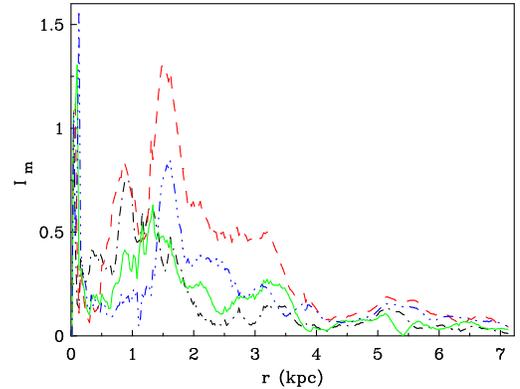}
 \end{center}
    \figcaption{Fourier power spectrum of the moment zero map shown in Figure \ref{fig-mom0}.  Modes up to $m=$4 are plotted as a function of radius with lines for $m=$1 in black dash-dot, $m=$2 in red dash, $m=$3 in green solid, and $m=$4 in blue dash-dot-dot-dot.
\label{fig-fourier}}
\end{figure}

\subsection{\label{sec:innerargs}Isolating the inner structure pattern speed}
As in previous applications of the method, we make use of a priori information to develop physically motivated models for $\Omega_p(r)$.  And as with all such models, in order that they supply rigorous estimates we must also account for evidence suggesting susceptibility to regularization-induced bias.  In the surface brightness (Figure \ref{fig-mom0}) and its Fourier decomposition (Figure \ref{fig-fourier}) we identify a region outside $r$$\sim$4 kpc, in particular, where both the surface brightness and power in the $m$=2 component are low, an indication that the information to be extracted there is potentially unreliable and difficult to constrain through modeling. \\
 \indent This is manifest in solutions for which regularization is employed throughout the full extent of the emission.  Bins at large radii in the lowest-$\chi^2$ solutions exhibit a significant degree of variation in their modeling, confirming that constraining the outer pattern speed is difficult.  According to the conclusion drawn by M08 in tests of the regularized TWR calculation on simulations, this challenges the accuracy with which all inner bins can realize the true pattern speed.  We therefore initially consider models which parameterize a cut radius $r_c$=4.1 kpc, beyond which all bins are calculated without regularization.  In testing, we find this cut radius to coincide with a clear minimum in the $\chi^2$, with all other best-fit parameters held fixed.  \\
\indent A second, shallower minimum at $r_c$=5.3 kpc is also compelling, and we consider its parameterization in models of $\Omega_p(r)$ in the analysis that follows, as well. This location may well be reasonable for the separation of the patterns given that it seems to match expectations for the location where the outer, material pattern begins.  We cite in particular the study of \citet{elmegreen1} who, like \citet{tully}, argue that OLR occurs at the termination of the bright, inner spiral structure and in the pretext of mode-coupling therein identify an overlap between the CR of the inner pattern with the ILR of a 10-20 $\kmskpc$ outer pattern (e.g. that first proposed for the material pattern by \citet{tully}).  This places the the innermost extent of the material pattern at $r$$\sim$6.0 kpc (adopted into the distance convention used here).  \citet{vog93} also argue for a similar corotation radius based on observations of streaming motions in the ionized gas component of the ISM.  And while this does not locate the inner extent of the outer pattern, it nevertheless implies that the outer arms are separate from, and have a lower pattern speed than, the inner arms \citep{vog93}.  Consequently, it is consistent with the conclusion of \citet{elmegreen1}.  \\
\indent A transition from an inner to an outer pattern is also recognizable in the tidal perturbation-only model of \citet{salob}.  There, an independent spiral pattern with corotation near $r$=4.6 kpc is found to be followed by structure at the lower 10-20 $\kmskpc$ pattern speed.  Any resonance overlap, however, they argue is likely coincidental since the value of the higher, inner speed is associated with the maximum in $\Omega$-$\kappa/2$, while the lower speed is determined mainly by external forcing.   \\
\indent In addition, as revealed in the sections to follow, while results with $r_c$=4.1 kpc indicate much higher speeds ($\sim$50-100 $\kmskpc$) than the TW method ($\Omega_p$=38 $\kmskpc$), with $r_c$=5.3 kpc solutions measure a much lower speed exterior to $r$$\simeq$4 kpc, at least qualitatively more consistent with the gross overall speed estimated with the TW calculation.
\subsection{\label{sec:inneronly}Best-fit models}
\begin{figure}
\begin{center}
 \leavevmode
\plotone{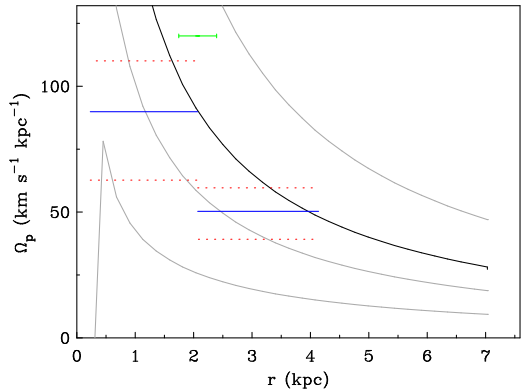}
\end{center}
\figcaption{
The best-fit regularized solution with $r_c$=4.1 kpc for PA=170$\degree$+/-5$\degree$.  For this solution, bins exterior to $r$=4.1 kpc (not shown) have been calculated without regularization.  Dashed red lines represent the difference from solutions derived with a two-pattern speed model at PA=165$\degree$ and 175$\degree$.  Horizontal error bars represent the dispersion in $r_{t,1}$ and $r_{t,2}$ from PA to PA.  The values in the zone of the bright spiral structure correspond to $\Omega_{p,1}$=90-27/+20 $\kmskpc$ out to $r_{t,1}$=2.1$\pm$0.3 kpc and $\Omega_{p,2}$=50+9/-11 $\kmskpc$ out to $r_c$.  Curves for $\Omega$, $\Omega$$\pm$$\kappa$/2 and $\Omega$-$\kappa$/4 (see $\S$ \ref{sec:corrollary}) are shown in gray. \label{fig-globalcut14a}}
\end{figure}
\begin{figure}
\begin{center}
 \leavevmode
\plotone{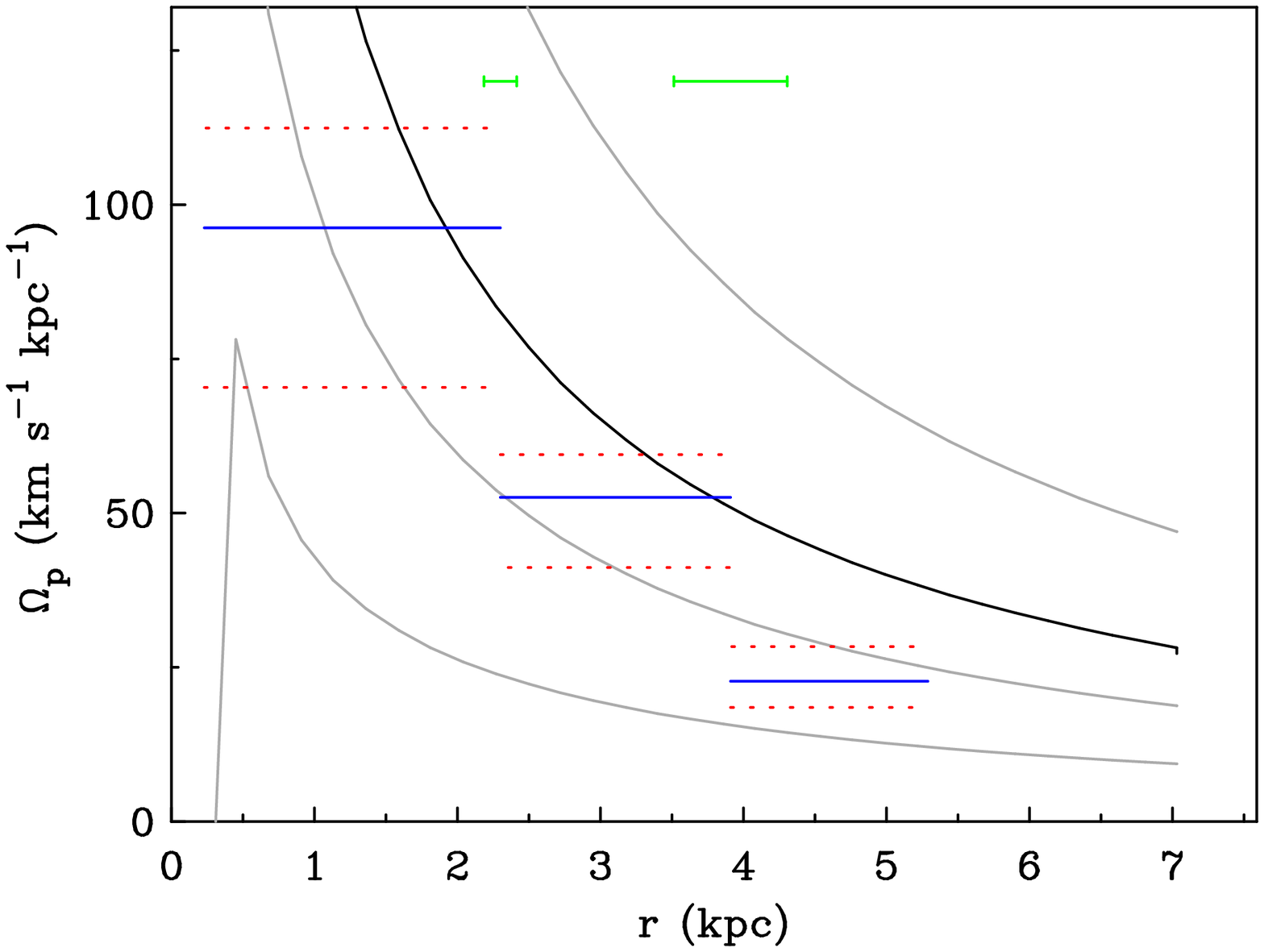}
\end{center}
\figcaption{The best-fit regularized solution with $r_c$=5.3 kpc for PA=170$\degree$+/-5$\degree$.  For this solution, bins exterior to $r$=5.3 kpc (not shown) have been calculated without regularization.  Dashed red lines represent the difference from solutions derived with a three-pattern speed model at PA=165$\degree$ and 175$\degree$.  Horizontal error bars represent the dispersion in $r_{t,1}$ and $r_{t,2}$ from PA to PA.  The values in the zone of the bright spiral structure correspond to $\Omega_{p,1}$=96-26/+16 $\kmskpc$ out to $r_{t,1}$=2.3$\pm$0.1 kpc, $\Omega_{p,2}$=51+7/-11 $\kmskpc$ out to $r_{t,2}$=3.9$\pm$0.4 kpc and $\Omega_{p,3}$=23-7/+6 $\kmskpc$ out to $r_c$.  Curves for $\Omega$, $\Omega$$\pm$$\kappa$/2 and $\Omega$-$\kappa$/4 (see $\S$ \ref{sec:corrollary}) are shown in gray. \label{fig-globalcut9a}}
\end{figure}
\begin{figure*}
\begin{center}
\epsscale{1.0}
\plottwo{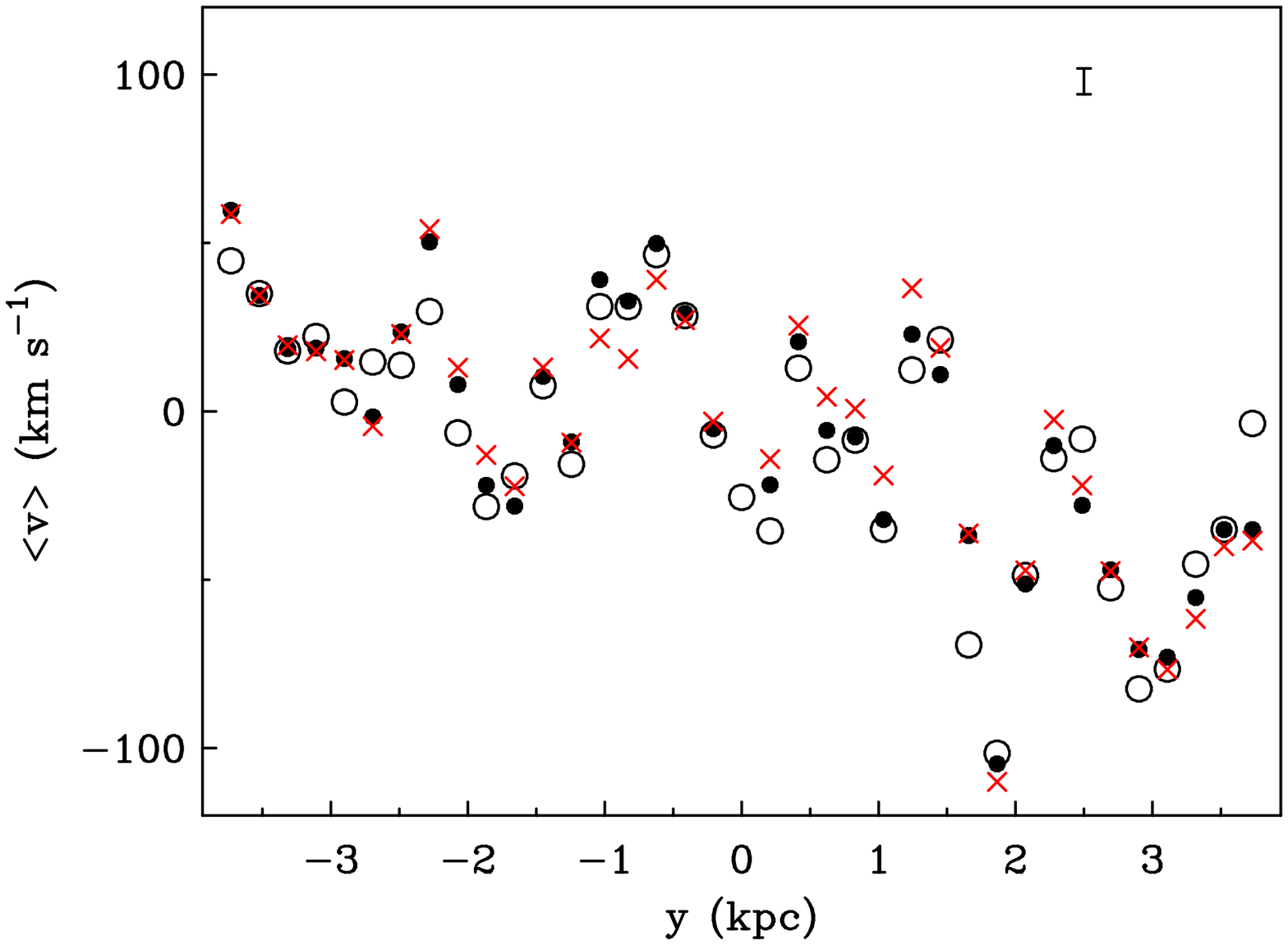}{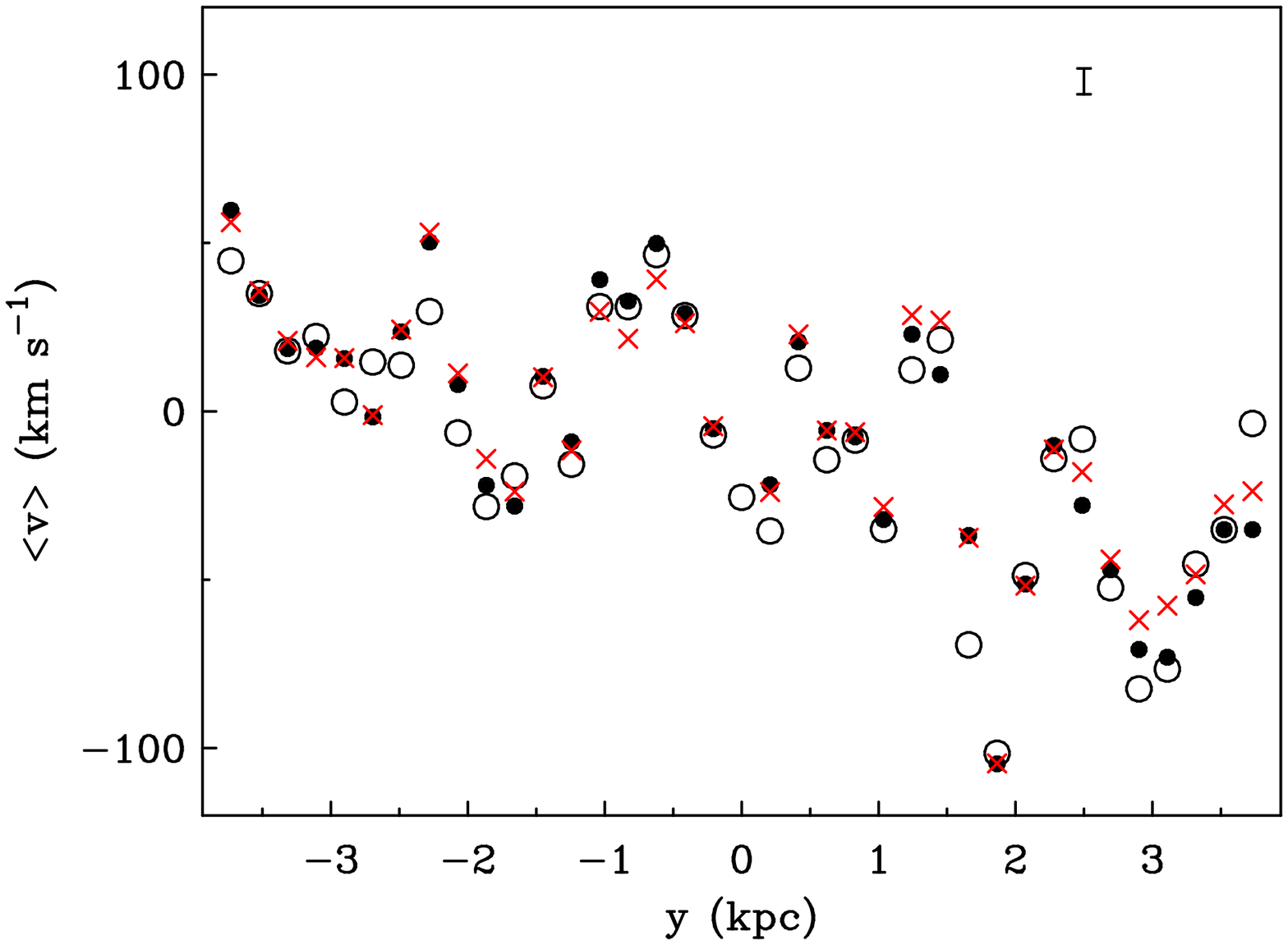}\\
\plottwo{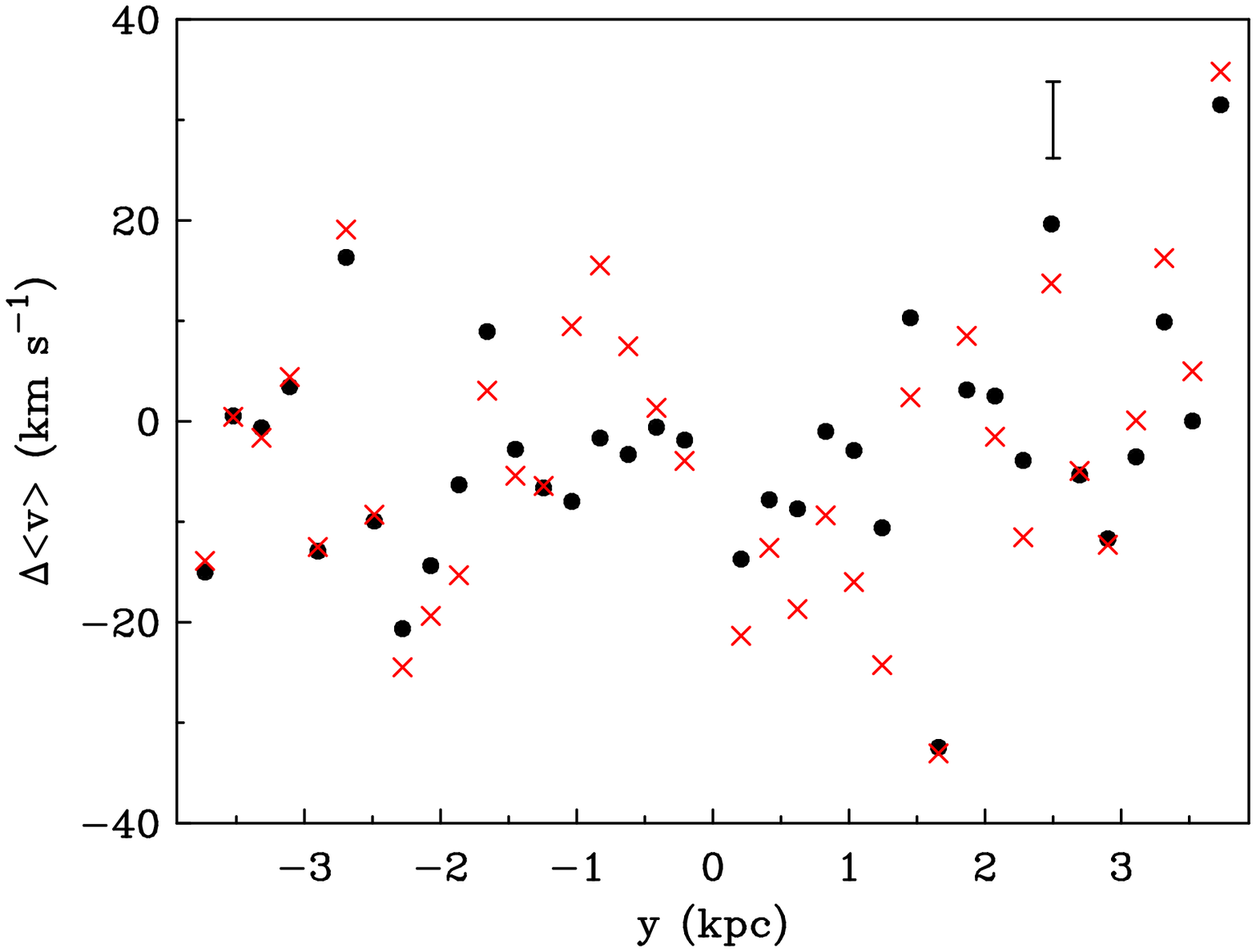}{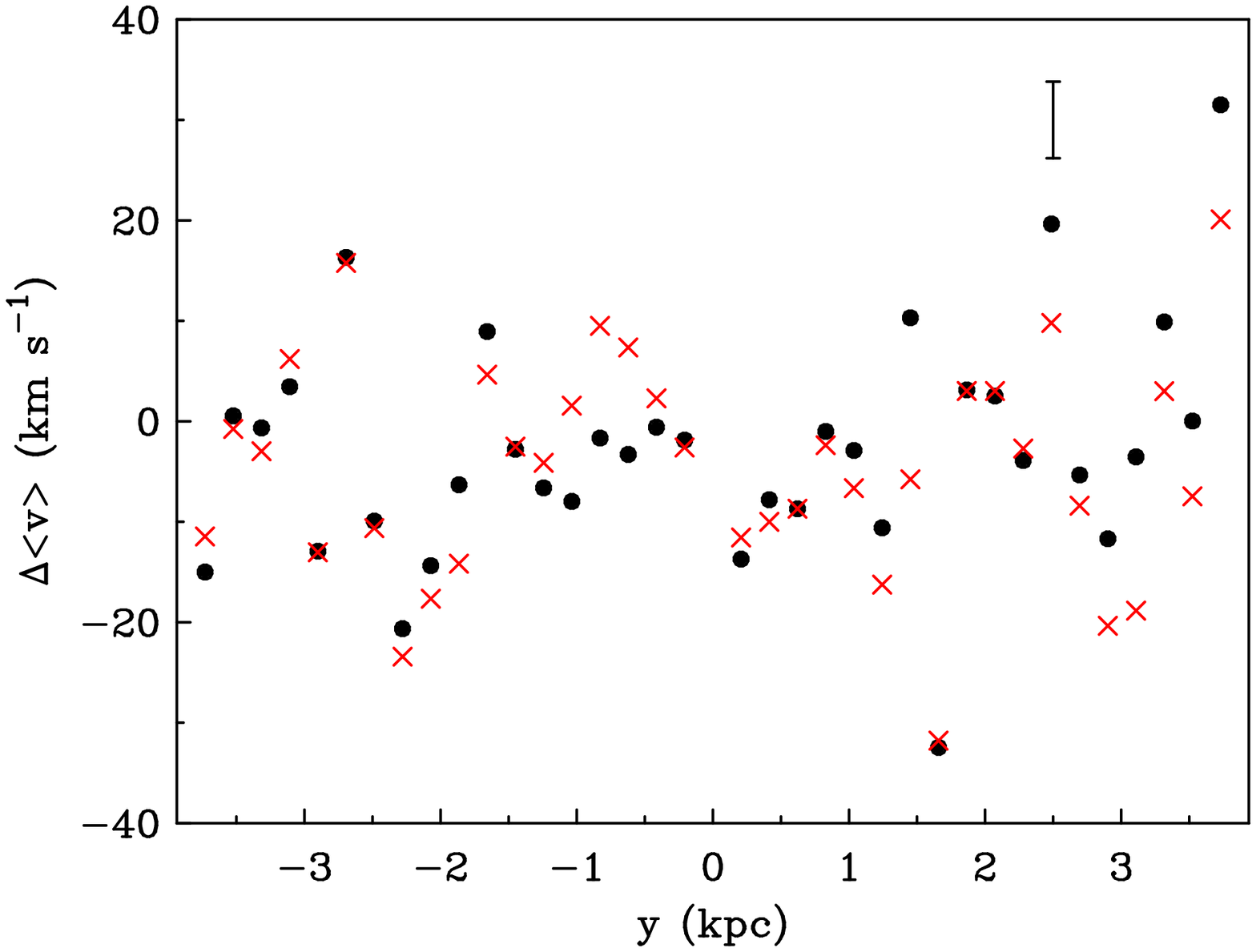}
 \end{center}
    \figcaption{(Top) Plots of solution-reproduced (dots and crosses) and actual (open circles) integrals $<$$v$$>_i$=$b_i$/$\int \Sigma dx$ as a function of slice position $y$ at PA=170$\degree$.  The values associated with the best-fit two-pattern speed solution calculated with $r_c$=4.1 kpc (black dots) are plotted along with those of the order-zero polynomial solution with constant pattern speed $\Omega_p$=55 $\kmskpc$ (red crosses; left panel) and the second order polynomial solution (red crosses; right panel).  (Bottom) Plots of the residuals in $<$$v$$>$ reproduced by the solutions considered above (left: black dots for the best-fit, red crosses for the constant solution; right: black dots for the best-fit, red crosses for the quadratic solution).  The adopted global error $\sigma^{<v>}$ is shown in the upper right in each plot.  Only those slices which show a contribution from bins inward of $r_c$=4.1 kpc are shown.  
\label{fig-vslice}}
\end{figure*}
\begin{deluxetable*}{crcccccccccccc}
\tablecolumns{14}

\tablewidth{0pt}
\tablecaption{$\chi^2$ Model Comparison\label{tab-chis}}
\tablehead{
\colhead{}&\colhead{} & \multicolumn{5}{c}{$r_c$=4.1 kpc} & \multicolumn{7}{c}{$r_c$=5.3 kpc} \\
\cline{1-14}
\colhead{PA}&\colhead{Model}& \multicolumn{2}{c}{$\Omega_p$}&\multicolumn{3}{c}{$\chi_{\nu,s}^2$}& \multicolumn{3}{c}{$\Omega_p$}&\multicolumn{4}{c}{$\chi_{\nu,s}^2$}}
\startdata
 & & $A$& $B$ &$A$& $B$& all& $A$& $B$& $C$&$A$& $B$& $C$& all\\
\cline{3-4}\cline{5-7}\cline{8-10}\cline{11-14}
& three speed& \nodata&\nodata &\nodata&\nodata &\nodata &96&  51&  23& 2.17 &2.61 & 2.56& 2.22  \\
PA=170$\degree$& two speed& 90&  50&1.1 &1.28& 1.65& 62&  62&  20&2.17 &2.61 & 2.56& 2.22 \\
& constant& 55& 55& 2.18 &1.58& 1.97& 27& 27& 27&  3.64 &2.72 &2.56& 2.86 \\
& quadratic& 95-&33&1.58 &2.52& 2.99&  180-&-&23&2.00 &4.18 &6.92& 3.76\\ 
\enddata
\tablecomments{Representation of the goodness of fit for model solutions with either $r_c$=4.1 kpc (left) or $r_c$=5.3 kpc (right) at the nominal PA=170$\degree$.  The $\chi^2$ estimates in the zones 0$<$$\vert y\vert$$<$2.3 kpc, 2.3$<$$\vert y\vert$$<$4.0 kpc, and 4.0$<$$\vert y\vert$$<$5.3 kpc, labeled $A$, $B$, and $C$, respectively, are listed along with those for all slices, where all values are calculated as a reduced $\chi^2$ difference of the model-reproduced to measured $<$$v$$>_i$.  The $\chi_{\nu,s}^2$ in each radial zone is normalized by the number of bins in that zone minus the number of degrees of freedom, and so represent a goodness-of-fit distinct from a $\chi_{\nu}^2$ over all slices with which we judge the whole solution.  The multi-speed solutions with $r_c$=4.1 and 5.3 kpc correspond to the best-fit two- and three-pattern speed solutions, respectively.  Pattern speed estimates in units of $\kmskpc$ in each zone are also listed. For quadratic solutions, the values in the first and last radial bins are indicated.}
\end{deluxetable*}

\indent When we minimize the influence of the suspected material pattern in the TWR solutions by calculating the bins in the outer zone without regularization, we in principle maximize the leverage on the inner structure.  In doing so, we find the data at the nominal PA to be well fit by two distinct pattern speeds interior to $r_c$=4.1 kpc. The overall pattern speed solution with PA uncertainty $\delta_{PA}$=$\pm$5$\degree$, to be discussed at length below, is represented in Figure \ref{fig-globalcut14a}.  Following the treatment of M08 for constructing errors on the measurement from a particular observational scenario, error bars represent the dispersion of the parameters in the best-fit solution derived with a two-pattern speed model at PA=165$\degree$, 170$\degree$ and 175$\degree$.  As will be discussed further in $\S$ \ref{sec:corrollary}, these two pattern speeds both end at corotation, within the uncertainties. \\
\indent To quantify the relative benefit of the two-pattern speed solution, in Table \ref{tab-chis} we list the $\chi_{\nu}^2$ estimate for several model solutions calculated over all slices in the TWR quadrature. In this table we also consider a $\chi_{\nu}^2$ over slices in the zones 0$<$$\vert y\vert$$<$2.3 kpc and 2.3$<$$\vert y\vert$$<$4.1 kpc at the nominal PA.  We expect the $<$$v$$>$ for slices in each of these zones to predominantly reflect measurements in the radial bins $r$=$\vert y\vert$, so these separated $\chi_{\nu}^2$ (labeled $\chi_{\nu,s}^2$ hereafter) should provide a fair comparison of $\Omega_p(r)$ from model to model at these radii. (Note, however, that all outer bins also appear in the $<$$v$$>$ reproduced by solutions in these zones.)  \\
\indent The $\chi_{\nu}^2$ fit over all slices principally suggests that the two pattern speed solution and the single, constant speed solution yield better agreement with the data than the quadratic solution.  For the former two solutions, the $\chi_{\nu}^2$ values are nearly indistinguishable at this PA (see the last column in Table \ref{tab-chis}). From the $\chi_{\nu,s}^2$, on the other hand, it is clear that inside $r$$\simeq$2 kpc the fit of the two pattern speed solution is significantly better than that of the constant speed.  \\
\indent Comparisons with the $<$$v$$>$ reproduced by the best-fit two-pattern speed model clearly demonstrate the incompatibility of the constant speed model, as shown on the left side of Figure \ref{fig-vslice}.  There, the $<$$v$$>_i$ (top) and residuals (bottom) at each slice position reproduced by the best-fit solution are plotted along with those reproduced by the best polynomial solution with constant pattern speed $\Omega_p$=55 $\kmskpc$ calculated over the same radial zone.  In the latter case, a greater departure from the measured values is readily apparent at slices inside $\vert y\vert$$\sim$2 kpc as compared with the best-fit solution, which transitions from an outer speed of $\Omega_p$=50 $\kmskpc$ to an inner speed of $\Omega_p$=90 $\kmskpc$ at $r_t$=2.3 kpc.  In fact, the constant solution fits the data better than the two-speed solution in only 2 of the 23 such slices.\\
\indent In constrast, the quadratic solution with $r_c$=4.1 kpc at PA=170$\degree$, which declines smoothly with radius from $\sim$95 $\kmskpc$, grants nearly comparable agreement with the measured values that the two pattern speeds entail, over a number of slices.  However, the fit of this solution weakens at slices between $\vert y\vert$$\sim$2.3-4 kpc (clear from the $\chi_{\nu,s}^2$), raising its $\chi_{\nu}^2$ well above that of the two-speed model. \\
\indent By comparison, then, it would seem that the two-pattern speed solution presents the best fit for slices at both small and large radii (in slices $\vert y\vert$$\leq$4 kpc). 
\subsubsection{\label{sec:rc53}Extended models}
\indent When the regularized zone 4.1$<$$r$$<$5.3 kpc is included in solutions, the best-fit solution once again measures two pattern speeds inside 4 kpc, but now a third, distinct pattern speed is also parameterized.  Figure \ref{fig-globalcut9a} plots this best-fit solution, where, again, error bars are defined by the dispersion in the lowest-$\chi_{\nu}^2$ solutions derived with a three-pattern speed model at PA=170$\degree$, 165$\degree$, and 175$\degree$.  \\
\indent According to the $\chi_{\nu,s}^2$ in Table \ref{tab-chis} for solutions with $r_c$=5.3 kpc, in both inner and outer zones this best-fit solution is superior to a two-speed solution which transitions from a single constant 62 $\kmskpc$ pattern speed inside 3.2 kpc to a lower 20 $\kmskpc$ speed and also, once again, to either of the two polynomial solutions considered here. \\
\indent The quadratic solution (decreasing from 180 $\kmskpc$) fits relatively well in the innermost bins, but overall the fit is now less comparable to that provided by solutions measuring three pattern speeds.  The agreement between the constant model solution-reproduced $<$$v$$>$ and the data also weakens, relative to the best-fit solution, especially inside $r$=4 kpc. This can be attributed to the decrease in the value measured with the constant model, from $\Omega_p$=55 $\kmskpc$ with $r_c$=4.1 kpc to $\Omega_p$=27 $\kmskpc$ with $r_c$=5.3 kpc.  Incidentally, this is a clear indication that not only is the pattern speed in the zone 4$\lesssim$$r$$\lesssim$5 kpc lower than $\Omega_p$=55 $\kmskpc$, but as such undeniably influences all bin values calculated inward with this type of model, thereby interfering with accurate measurement interior.  \\
\indent In contrast, when the zone 4$\lesssim$$r$$<$5.3 kpc is distinct and isolated, multi-speed models are nearly free of such inaccuracy.  In both the two- and three-speed solutions with $r_c$=5.3 kpc, the inner pattern speeds are nearly identical to the values measured in solutions with $r_c$=4.1 kpc. This seems to suggest that, despite the evidence in Figures \ref{fig-mom0} and \ref{fig-fourier} that information beyond $r$$\simeq$4 kpc is not conducive to modeling and extraction, the determination in the third zone is fairly accurate.  \\
\indent This equivalence inside $r$$\simeq$4 kpc to the pattern speeds measured in the solutions with $r_c$=4.1 kpc derives in practice through the parameterization of a transition at $r$$\simeq$4 kpc.  This establishes an identical radial domain for the inner speeds in the solutions with $r_c$=5.3 kpc and 4.1 kpc.  The transition $r_{t,1}$$\sim$2.3 kpc in the three-speed solution as such yields the greatest similarity to the best-fit two-speed solution with $r_c$=4.1 kpc.  \\
\indent Moreover, according to Equation \ref{eq:twr}, rigorous measurement inside 4 kpc in principle also owes to accurate measurement for the pattern speed in the zone 4$\lesssim$$r$$\lesssim$5 kpc.  As inferred above, the measurement $\Omega_{p,3}$=20 $\kmskpc$ is in fact lower than all measurements interior.  Presumably, it is the value in this zone that contributes to the measurement of the rather low TW value 38 $\kmskpc$ \citep{zrm04}.\\
\indent Even if the measurement for a distinct speed in the zone 4$\lesssim$$r$$\lesssim$5 kpc is a good description of the pattern there, since all $\chi^2$ are lower in solutions with $r_c$=4.1 kpc than with $r_c$=5.3 kpc, we take this as an indication that calculating bins in the zone 4.1 $<$$r$$<$5.3 kpc without regularization does not reintroduce noise into solutions. Consequently, solutions with $r_c$=4.1 kpc should yield the more accurate description for structure in the zone $r$$\lesssim$4 kpc.  This analysis therefore at best indicates that within $r_c$=4.1 kpc the data at the nominal PA are well fit by two pattern speeds.  In addition, though, it seems possible to extend the multi-speed model's estimate for $\Omega_p(r)$ to 5.3 kpc without loss of validity, and this appears to be a good approximation to the pattern speeds of the structure across this zone.  \\ 
\indent Future high-resolution CARMA observations of M51 should enable the TWR measurements inside $r$$\lesssim$4 kpc to be more clearly distinguished, especially at the innermost radial bins.  Presently, however, it is nevertheless clear that with the radial calculation at PA=170$\degree$ we measure a pattern speed for the bright spiral structure in the zone $r$$\lesssim$4 kpc higher than the global $\sim$38 $\kmskpc$ found by \citet{zrm04}, also at PA=170$\degree$.  Interestingly, our measurement of a higher inner speed resembles the lower bound on such a pattern available with the TW calculation, $\Omega_p$$\geq$88 $\kmskpc$ \citep{zrm04}. 
\subsection{\label{sec:PAchar}Dependence on PA}
\begin{figure}
\begin{center}
 \leavevmode
\plotone{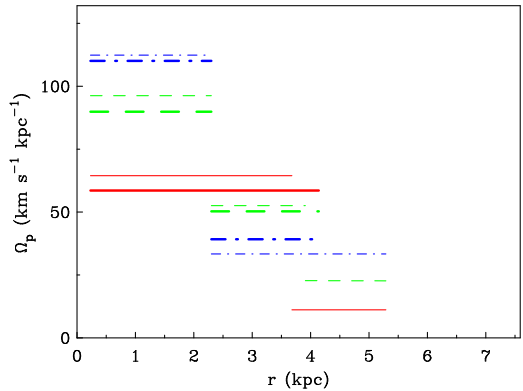}
\end{center}
\figcaption{Best-fit regularized solutions at three PAs (PA=165$\degree$, 170$\degree$, and 175$\degree$).  Solutions calculated with both $r_c$=4.1 kpc and 5.3 kpc are plotted, with the latter shown in thinner line.  In these solutions, bins exterior to $r_c$ (not shown) are calculated without regularization. The solutions from PA=165$\degree$, 170$\degree$, and 175$\degree$ are shown in blue dash-dot, green dash and red solid, respectively.  Values and domains of the pattern speeds in solutions at each PA are given in Table \ref{tab-chisc14c9165175}. 
\label{fig-all3}}
\end{figure}
\begin{deluxetable*}{crcccccccccccc}
\tablecolumns{14}
\tablewidth{0pt}
\tablecaption{$\chi^2$ Model Comparisons\label{tab-chisc14c9165175}}
\tablehead{
\colhead{}&\colhead{} & \multicolumn{5}{c}{$r_c$=4.1 kpc} & \multicolumn{7}{c}{$r_c$=5.3 kpc}\\
\cline{1-14}
\colhead{PA}&\colhead{Model}& \multicolumn{2}{c}{$\Omega_p$}&\multicolumn{3}{c}{$\chi_{\nu,s}^2$}& \multicolumn{3}{c}{$\Omega_p$}&\multicolumn{4}{c}{$\chi_{\nu,s}^2$}}
\startdata
 & & $A$& $B$ &$A$& $B$& all& $A$& $B$& $C$&$A$& $B$& $C$& all\\
\cline{3-4}\cline{5-7}\cline{8-10}\cline{11-14}
& three speed& \nodata&\nodata &\nodata&\nodata &\nodata &112&  41&  28& 0.58 &0.84 &0.60& 1.70 \\
PA=165$\degree$& two speed& 110&  39&0.7 &0.64& 1.51& 112&  33&  33&0.56 &0.82 & 0.68& 1.66\\
& constant&  42& 42& 3.68 &0.76& 2.42 & 35& 35& 35&  3.26 &0.80 &1.02& 2.65 \\
& quadratic& 199-&16&2.58 &1.44& 3.22& 200-&-&30&1.74 &1.54 &1.18& 3.31\\ 

\cline{1-14}
& three speed& \nodata&\nodata &\nodata&\nodata &\nodata & 70& 59& 18&1.56 &3.18 &9.62& 3.26 \\
PA=175$\degree$& two speed& 63&  59& 1.66 &2.66& 1.81 &  64& 64& 11& 3.06 &5.22 &6.58& 2.82 \\
& constant& 59&  59&1.53 &2.28& 1.66 & 19& 19& 19& 3.96 &5.14 &6.96& 3.55\\
& quadratic& 73-&38&2.41 &4.12& 2.97&  109-&-&12& 3.22 &7.1 &8.6& 4.22\\ 
\enddata
\tablecomments{Representation of the goodness of fit given by $\chi_{\nu,s}^2$, as in Table \ref{tab-chis}, for model solutions at PA=165$\degree$ and 175$\degree$ calculated with $r_c$=4.1 kpc (left) and $r_c$=5.3 kpc (right). Here, the zones 0$<$$\vert y\vert$$<$$r_{t,1}$, $r_{t,1}$$<$$\vert y\vert$$<$$r_{t,2}$, and $r_{t,2}$$<$$\vert y\vert$$<$5.3 kpc are labeled $A$, $B$, and $C$, respectively, with transition radii as identified in the lowest-$\chi^2$ two (three) pattern speed solution calculated with $r_c$=4.1 kpc (5.3 kpc).  At 165$\degree$, these transitions occur at $r_{t,1}$=$r_t$=2.3 kpc where $r_c$=4.1 kpc and $r_{t,1}$=2.3 kpc and $r_{t,2}$=4.4 kpc where $r_c$=5.3 kpc.  At 175$\degree$, $r_{t,1}$=$r_t$=2.3 kpc where $r_c$=4.1 kpc and $r_{t,1}$=2.3 kpc and $r_{t,2}$=3.7 kpc where $r_c$=5.3 kpc. Pattern speed estimates in units of $\kmskpc$ in each zone are also listed. For quadratic solutions, the values in the first and last radial bins are indicated.}
\end{deluxetable*}
\begin{figure}
\begin{center}
 \leavevmode
\plotone{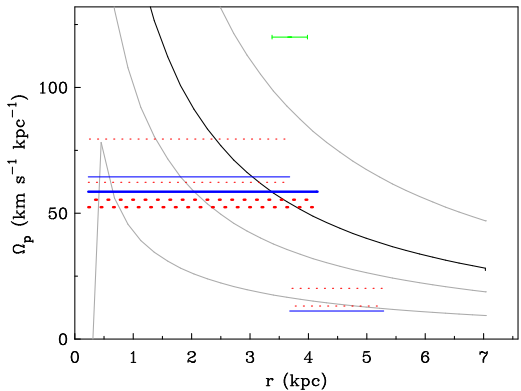}
\end{center}
\figcaption{The best-fit regularized solution for PA=175$\degree$+/-5$\degree$ and $r_c$=4.1 kpc (5.3 kpc).  Dashed red lines represent the difference from solutions derived with a one- (two-)pattern speed model at PA=170$\degree$ and 180$\degree$ (the best fit functional forms at PA=175$\degree$).  Horizontal error bars represent the dispersion in $r_{t}$ calculated in solutions with $r_c$=5.3 kpc from PA to PA.  Where $r_c$=4.1 kpc, the value in the zone of the bright spiral structure corresponds to $\Omega_{p,1}$=59-7/-3 $\kmskpc$ out to $r_c$=4.1 kpc and where $r_c$=5.3 kpc (shown in thinner line), $\Omega_{p,1}$=64-2/+15 $\kmskpc$ out to $r_t$=3.7$\pm$0.2 and $\Omega_{p,2}$=11+2/+9 $\kmskpc$ out to $r_c$=5.3 kpc.  Curves for $\Omega$, $\Omega$$\pm$$\kappa$/2 and $\Omega$-$\kappa$/4 (see $\S$ \ref{sec:corrollary}) are shown in gray. \label{fig-twfit175bestc14}}
\end{figure}
\indent We expect the rather large PA uncertainty $\delta_{PA}$=$\pm$5$\degree$ to introduce significant variation in the values measured at the nominal PA; in the previous section, we used this to define the error in the measurement of the best-fit parameters for PA=170$\degree$.  But TWR solutions from PA=165$\degree$ and 175$\degree$ themselves additionally indicate a departure from the parametrization characteristic of the lowest $\chi_{\nu}^2$ solution measured at 170$\degree$.  \\
\indent When we identify the best-fit solutions strictly by their $\chi_{\nu}^2$ over all slices {\it at each PA}--rather than restrict our consideration at PA=165$\degree$ and 175$\degree$ to pattern speed solutions optimal at PA=170$\degree$--we find that the values and domains of the best-fit pattern speeds vary from PA to PA.  Figure \ref{fig-all3} shows the best-fit solutions at the three PAs, the values and $\chi_{\nu}^2$ for which are given in Table \ref{tab-chisc14c9165175}.  There, solutions with $r_c$=5.3 kpc at PA=165$\degree$ and 175$\degree$, unlike at PA=170$\degree$, measure at most two distinct pattern speeds.  More notably, although the best-fit solution with $r_c$=4.1 kpc at 165$\degree$ measures two pattern speeds, at 175$\degree$ no unique pattern speed is measured inside $r$$\lesssim$2 kpc.  \\
\indent Model comparisons based on $\chi_{\nu}^2$ over all slices and the separated $\chi_{\nu,s}^2$ diagnostic (Table \ref{tab-chisc14c9165175}) demonstrate the degree to which these best-fit solutions differ from those at the nominal PA.  At 165$\degree$, for instance, the $\chi_{\nu}^2$ for solutions with $r_c$=5.3 kpc suggest that three pattern speeds are nearly indistinguishable from the best-fit solution.  The $\chi_{\nu,s}^2$ confirms that the third zone in solutions with $r_c$=5.3 kpc is fit equally as well by a third pattern speed as by the second speed in Figure \ref{fig-all3}.  (Inside $r$$\sim$4 kpc, the $\chi^2$ (and speeds) of the three and two speed solutions are nearly identical.)  Furthermore, where $r_c$=4.1 kpc two pattern speeds fit the data significantly better than a single, constant pattern speed. \\ 
\indent At PA=175$\degree$, too, judged overall by the $\chi_{\nu}^2$, three (two) distinct pattern speeds seem to fit nearly as well as the best-fit solution with $r_c$=5.3 kpc (4.1 kpc).  From the $\chi_{\nu,s}^2$ it is apparent that, for $r_c$=5.3 kpc two pattern speeds inside $r$$\simeq$4 kpc fit the inner two zones significantly better than the single pattern speed shown in Figure \ref{fig-all3}.  However, the $\chi_{\nu,s}^2$ in the third zone of this triple pattern speed solution is rather high; the small 11 $\kmskpc$ difference in the speeds measured inside $r$$\sim$4 kpc therefore seems available only at the expense of accuracy in third pattern speed.  Furthermore, for $r_c$=4.1 kpc the distinction between the two pattern speeds measured inside $r$$\simeq$4 kpc weakens; the two in this case are nearly identical to the constant value (such that the $\chi_{\nu}^2 $ (and $\chi_{\nu,s}^2$) of both solutions are comparable).  Overall, then, Table \ref{tab-chisc14c9165175} supports a conclusion that the data at 175$\degree$ are at best consistent with a single constant pattern speed inside $r$$\sim$4 kpc.  \\
\indent This apparent preference for the measurement of a constant pattern speed at PA=175$\degree$ may suggest a phenomenological difference in the projection of asymmetries (both intensity and velocity) from that in the 170$\degree$ case.  For example, at PA=175$\degree$, and also at a higher PA=180$\degree$, $<$$v$$>$ measurements in slices $\vert y\vert$$<$2.3 kpc are about $\sim$50 \% smaller than at PA=170$\degree$, a significant difference given the flux error for these slices.  While this is consistent with expectations for the large change in TW integrals introduced by a change in the PA (e.g. from a nominal $\delta_{PA}$=0$\degree$; \citealt{debPA}), the smaller projected streaming velocities in this case may be more difficult to extract with the TWR calculation, and lacking strong signatures, reproducing the higher pattern speed measured at 165$\degree$ and 170$\degree$ may therefore be improbable at 175$\degree$.  In effect, the single measured speed inside $r$$\simeq$4 kpc at PA=175$\degree$ may therefore reasonably describe two pattern speeds with significant error in each; this speed is nearly consistent with the solution plotted in Figure \ref{fig-globalcut14a} with the errors defined by the PA uncertainty.  \\
\indent Critically, however, the data admit both PA=170$\degree$ and PA=175$\degree$, and according to the findings of \citet{shetty} addressed in section \ref{sec:varyPA}, the latter may be arguably more valid at the inner radii than our chosen 170$\degree$.  In this case, if, as might be indicated by the analysis of \citet{shetty}, the PA does not reach the assumed 170$\degree$ until $r$$\simeq$3 kpc, rather than measuring a distinct pattern speed, $\Omega_{p,1}$ could thus be interpreted as simply identifying the region in the disk where the assumed PA is inappropriate.  Furthermore, by the same token as above, the large (by comparison) $<$$v$$>$ at 170$\degree$ may themselves reflect a misrepresentation of velocities in projection from the 175$\degree$ case.  Consequently, if choosing 170$\degree$ introduces streaming motions that are unreal, the two pattern speeds measured in the best-fit solution at 170$\degree$ could just as persuasively reflect a large PA error introduced into the measure of a single constant pattern speed.  \\
\indent Since solutions at the two PAs indicate quite independent radial behaviors, we include here an estimate which may be more appropriate for a nominal PA=175$\degree$.  Figure \ref{fig-twfit175bestc14} plots solutions with $r_c$=4.1 kpc and $r_c$=5.3 kpc for PA=175$\degree$+/-5$\degree$ where we have fixed the functional form of solutions at PA=170$\degree$ and 180$\degree$ to that of the best-fit 175$\degree$ solution.  (These solutions are also the best-fitting at PA=180$\degree$; for 170$\degree$ the multi-pattern speed solutions in the previous section are otherwise best).  Note that the values inside $r$$\simeq$4 kpc are only slightly modified with the inclusion of the zone 4.1$<$$r$$<$5.3 kpc. 
\subsection{\label{sec:state}The state of current measurements}
\indent Although the PA of the disk is ambiguous, for the particular case of a single assumed PA=170$\degree$, the majority of our analysis leads us to consider the solution in Figure \ref{fig-globalcut14a} a fair representation of the (isolated) inner disk.  As stated previously, the errors represent PA uncertainty introduced to the parameters of the best-fit solution derived at the nominal PA with a two-pattern speed model.  This uncertainty $\delta_{PA}$=$\pm$5$\degree$ defines 22\% and 16\% error on the pattern speed estimates $\Omega_{p,1}$ and $\Omega_{p,2}$, respectively, and 14\% error in the transition $r_{t}$, all reasonable with regard to the standard set by the study of \citet{debPA}.  \\
\indent According to the study of \citet{meidt}, part of this error can be expected to have originated with limitations in determining the location of the transition between the two patterns (assuming they exist), as a result of the finite radial bin width.  In addition, for the inner pattern speed additional uncertainty may arise given the disparity between the inner extent of the solution and that of the true, dominant two-armed pattern, which in the surface brightness terminates at the ring-like structure at $r$$\sim$0.6 kpc.  If structure inside $r$$\simeq$1 kpc, perhaps like that identified in the near-IR by \citet{zrr}, contributes to the calculation with a unique pattern speed in this zone, our measurement $\Omega_{p,1}$ would represent a combination of this value with that for structure out to $r$$\sim$2 kpc.  (Note, too, in this case, $\Omega_{p,1}$ would also mis-estimate the true pattern speed between 0.6$\lesssim$$r$$\lesssim$2.0 kpc.)  Unfortunately, identifying whether or not an additional, unique pattern exists inside $r$$\sim$1.0 kpc, or even establishing an innermost extent for the measure $\Omega_{p,1}$ is currently beyond our capability; the total degrees of freedom for even the lowest order polynomial exceed the number of available bins in the innermost radial zone.  \\ 
\indent Presently, the pattern speeds in the best-fit solution at PA=170$\degree$ in general tend to be arranged adjacent to the angular rotation curve (or perhaps even along the curve $\Omega-\kappa/4$; see Figure \ref{fig-globalcut14a} or Figure \ref{fig-globalcut9a} showing rotation curves established in $\S$ \ref{sec:corrollary}), much as if identifying a propensity towards a material pattern description.  Rather than furnish a description for arms that are material and winding, however, we note that the very alignment of multiple pattern speeds with the disk angular rotation may relate to an underlying mechanism governing the existence and maintenance of structure in the disk.  In one interpretation, the succession of corotation radii implied by the best-fit solution might be an indication of resonance overlap, as discussed inconclusively in $\S$ \ref{sec:corrollary}.  Associated with mode-coupling, this would allow quasi-static spiral structure to be maintained over a large portion of the disk \citep{syg} while transporting energy and angular momentum outward.  \\
\indent Our TWR solutions furthermore seem unlike what might be expected for transient density waves, with description deriving from the propagation of tidal perturbations studied by \citealt{salob}. For example, although their range of applicability seems limited to the innermost radii (but taken as an approximation to the best-fit multi-speed solutions) the bin values in the current set of quadratic solutions are much closer to the angular rotation of the disk than $\Omega-\kappa/2$, near which much of the $m$=2 structure in the models of \citet{salob} achieves its greatest amplitude.  \\
\indent In order to best establish the extent to which the TWR solution in Figure \ref{fig-globalcut14a} is truly a valid description of the bright spiral structure, in the section immediately following, and in $\S\S$ \ref{sec:varyPA} to \ref{sec:discussion}, we relate the radial dependence exhibited by the solution to observed morphological and kinematic structure.  The inner disk of M51 has been suggested to sustain radial variation in the PA \citep{shetty} and an additional $m$=3 mode \citep{henry}, both of which undeniably challenge the authenticity of the TWR solutions, and so we address the possible influence of each of these in turn. 
\subsection{\label{sec:corrollary}Possible complimentary evidence for multiple pattern speeds and indications of mode coupling}
\indent Though perhaps unexpected, the identification of a transition between two pattern speeds in the inner disk seems supported by independent studies of the bright spiral structure.  At least two sections best fit with slightly different pitch angles $i_p$ have been identified in both spiral arms, possibly the signature of two or more distinct pattern speeds.  Notably, the anisotropic wavelet approach of \citet{patrikeev} shows evidence for extreme departures from the conventionally adopted value $i_p$=21$\degree$.  The maximum occurring nearly symmetrically in both arms at $r$$\sim$2 kpc (see Figures 6, 7a and 8a in \citealt{patrikeev})--very near the transition $r_t$ identified in our best-fit solution at PA=170$\degree$--is especially compelling since it may indicate more than a simple departure from a logarithmic dependence.  The transition $r_{t,2}$ in Figure \ref{fig-globalcut9a} also occurs near a maximum in $i_p$. (To be sure, the other extrema in $i_p$ imply no such correlation).  This analysis is largely consistent with the findings of \citet{henry} covering radii $r$$\lesssim$4 kpc which identify three arm sections each with a unique $i_p$.\\
\indent Of course, a systematic error in deprojection, such as due to an incorrectly adopted PA or inclination angle (or a radially varying PA, as considered in $\S$ \ref{sec:varyPA}) could very well alone produce the effect measured by \citet{patrikeev} (where PA=170$\degree$). A firm conclusion might therefore require a better understanding of how a change in pitch angle at a given radius relates to a change in pattern speed, for instance (assuming that the spirals are indeed logarithmic). \\
\indent The transition between two patterns inside $r\simeq$4 kpc indicated in our solution also seems significant given that it coincides with features in the zeroth moment map's Fourier decomposition (Figure \ref{fig-fourier}); as at $r$$\sim$4 kpc, the power in the $m$=2 mode is characterized by a decline at r$\sim$2 kpc possibly marking the termination of a distinct structure.  (The same can be inferred at 4 kpc which coincides with the transition $r_{t,2}$ in the three-pattern speed solution.)\\
\indent Perhaps more compellingly, the transitions parameterized in our solutions at 170$\degree$ also appear to coincide with resonances, as illustrated in Figures \ref{fig-globalcut14a} and \ref{fig-globalcut9a}.  This, of course, would seem to depend largely on the assumed rotation curve.  As first demonstrated by \citet{tully}, streaming motions appear non-negligibly in the rotation curve of M51, making the true circular velocity difficult to constrain.  As more recently cataloged by \citet{shetty}, all other kinematic parameters are likewise susceptible to such errors, and so rotation curves generated with them are prone to inaccuracy.  To reduce the impact of streaming motions (and perhaps other systematic errors) on our resonance identifications, we fit our own ROTCUR-derived rotation curve with the commonly used approximation (e.g. by \citealt{fabgall})  
\begin{equation}
V_{rot}(r)=\frac{V_{max}(r/r_{max})}{\left(1/3+2/3(r/r_{max})^n\right)^{3/2n}}
\end{equation}
which yields a smoothed curve for $\Omega$.  In this expression, $V_{max}$ is the maximum rotational velocity, $r_{max}$ is the location where $V_{max}$ occurs, and $n$ determines how rapidly the curve becomes Keplerian.  Alternative fits (like that used on the inner 30'' by \citealt{aalto}) supply similar conclusions.\\
\indent With the resulting curves for $\Omega$, $\Omega-\kappa/2$, $\Omega-\kappa/4$, and $\Omega+\kappa/2$ plotted in Figures \ref{fig-globalcut14a} and \ref{fig-globalcut9a} we highlight the possible locations for the corotation, inner Lindblad, inner 4:1 ultraharmonic, and outer Lindblad resonances (or CR, ILR, UHR, and OLR) for each measured pattern speed.  Immediately we notice that both pattern speeds in the solution with $r_c$=4.1 kpc end at their CR within the uncertainties.  This circumstance is consistent with an early prediction for where spirals terminate (i.e. \citealt{lin}), which later yielded to findings that spirals can extend as far as OLR, if sometimes faintly (see \citealt{elmegreen1}, for example).  \\
\indent In addition, the transition between the two pattern speeds appears to occur at a resonance overlap.  As demonstrated in Figure \ref{fig-globalcut14a}, the CR of $\Omega_{p,1}$ overlaps the UHR of the pattern with $\Omega_{p,2}$.  Such coincidences have been identified in barred spiral simulations of \citet{rs} and \citet{debetal}.  As the former investigate, this overlap at resonance may be characteristic of non-linear mode coupling (e.g. \citealt{tagger} and \citealt{syg}) whereby energy and angular momentum are transferred between the modes.  But in contrast to the CR-ILR overlaps studied by \citet{masset}, which are accompanied by boosted beat modes detectable in the simulation power spectra at the overlap, they find no comparable evidence for mode-coupling in the case of the CR-UHR overlap.  (They suggest this overlap may nevertheless be related to a physical process.)\\
\indent  A CR-ILR overlap between $\Omega_{p,1}$ and $\Omega_{p,3}$ in the solution with $r_c$=5.3 kpc, on the otherhand, may be viable within the uncertainties.  However, between $\Omega_{p,2}$ and $\Omega_{p,3}$ in Figure \ref{fig-globalcut9a} a similar resonance overlap is not so clear; near the transition $r_{t,2}$ CR of $\Omega_{p,2}$ falls between the ILR and the UHR of $\Omega_{p,3}$. \\
\indent Figure \ref{fig-globalcut14a} also exhibits a turnover in the curve $\Omega-\kappa/2$, suggesting that patterns with angular speeds above the maximum lack an ILR.  However, given uncertainty in the rotation curve, this is difficult to constrain: while the angular frequency curves of \citet{tully} indicate that $\Omega-\kappa/2$$\sim$47 $\kmskpc$ at the turnover (also reproduced in the \citet{salob} model), we find that the maximum occurs at $\sim$75 $\kmskpc$.  (Our fit for $\Omega(r)$ may be slightly steep inside $\sim$1.0 kpc.)  Nevertheless, it is apparent for this solution that $\Omega_{p,1}$ lacks an ILR.  This suggests that a (trailing) wave with $\Omega_{p,1}$ can reflect from the center as a leading structure, a circumstance complimentary to the \citet{scoville} HST observations of central leading waves, as pointed out by \citet{salob}.  
\subsection{\label{sec:varyPA}Effect of a radial variation in PA}

From their analysis of the CO and H$_{\alpha}$ kinematics, \citet{shetty} find evidence for large non-zero radial flux (as measured by the mass/surface brightness-weighted radial velocity) in radial ranges that depend on the choice of PA.  From Figure 18 of that study, in particular, \citet{shetty} speculate that mass flux could be conserved should the disk of M51 sustain a radially-dependent PA (and/or inclination). This could be approximately achieved with PA=180$\degree$ out to $r$$\sim$1.8 kpc, PA=175$\degree$ out to $r$$\sim$2.8 kpc and PA=170$\degree$ out to $r$$\sim$3.7 kpc.  (The inclination angle, which might also be expected to vary, is much harder to account for in the TWR calculation.) \\
\indent If the PA does vary radially then the measurement in Figure \ref{fig-globalcut14a} (or Figure \ref{fig-twfit175bestc14}) could be affected by projection errors at certain radii.  Note that a radially varying PA implies a warp in the inner disk (which, if real, could be due to the presence of the companion) and so the disk would also not meet the assumptions of the TWR method.  Interpreted in this manner, our finding of possible multiple pattern speeds in the inner disk may be the result of such an effect.  \\
\begin{figure}
\begin{center}
\plotone{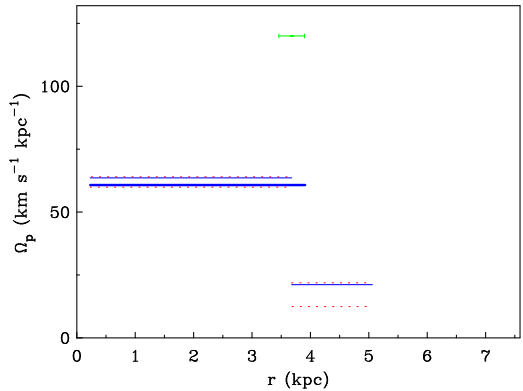}
 \end{center}
    \figcaption{The best-fit regularized solution for the PA twist map with $\delta_{PA}$=$\pm$2$\degree$ and $r_c$=4.1 kpc (5.3 kpc) shown in thick (thin) line.  (See text for a description.)  Dashed red lines represent the difference from solutions derived with a two-pattern speed model best for the $\delta_{PA}$=0$\degree$ case; errors for the solution with $r_c$=4.1 kpc are nearly coincident with these lines and have been left off for clarity.  Horizontal error bars represent the dispersion in $r_{t}$ calculated in solutions with $r_c$=5.3 kpc.  Where $r_c$=4.1 kpc, the value in the zone of the bright spiral structure corresponds to $\Omega_{p}^{PA}$=61+4/-1$\kmskpc$ out to $r_c$=4.1 kpc and where $r_c$=5.3 kpc (shown in thinner line), $\Omega_{p,1}^{PA}$=62$\pm$2 $\kmskpc$ out to $r_t$=3.8$\pm$0.5 kpc and $\Omega_{p,2}^{PA}$=18$\pm$2 $\kmskpc$ out to $r_c$=5.3 kpc.  For there solutions, bins exterior to $r_c$ (not shown) are calculated without regularization.  
\label{fig-twrgridb2}}
\end{figure}
We explore this possibility by allowing the PA to vary radially in the TWR quadrature according to the prescription given at the beginning of this subsection.  For simplicity, we retain $i$=24$\degree$ throughout the disk and let PA=170$\degree$ at all radii beyond $r$$\sim$3.7 kpc.  Figure \ref{fig-twrgridb2} shows the best-fit solutions with $r_c$=4.1 kpc.  Errors represent a residual PA uncertainty $\delta_{PA}$=$\pm$2$\degree$ estimated from Figure 18 of \citet{shetty}.  \\
\indent Interestingly, the global pattern speed inside $r_c$=4.1 kpc ($\Omega_{p}^{PA}$=62$\pm$2 $\kmskpc$) closely resembles the measurement at PA=175$\degree$ (Figure \ref{fig-twfit175bestc14}). The best-fit solution with $r_c$=5.3 kpc ($\Omega_{p,1}^{PA}$=62$\pm$2 $\kmskpc$ out to $r_t$=3.8$\pm$0.5 kpc, $\Omega_{p,2}^{PA}$=18$\pm$2 $\kmskpc$ out to $r_c$) also resembles that at 175$\degree$, and here the pattern speed in the zone 4$\lesssim$$r$$\lesssim$5 kpc does not seem to be the result of an incorrectly assumed PA (i.e. PA=170$\degree$ instead of 165$\degree$); extending the twist by another 5$\degree$ at radii greater than $r$=3.7 kpc produces little change in the calculated solutions.  \\
\indent That the estimates in Figures \ref{fig-twrgridb2} and \ref{fig-twfit175bestc14} are so similar seems to suggest the twist solution is less a manifestation of the PA twist than an indication that the PA assumed here inside $r$$\simeq$3 kpc is everywhere closer to 175$\degree$ than 170$\degree$.  The mean PA of the twist is 175$\degree$, so it may be reasonable to infer that the twist solution predominantly reflects information nearly identically to the 175$\degree$ case.  Solutions at 180$\degree$, too, are very similar to those at 175$\degree$, as indicated by the estimate assembled in Figure \ref{fig-twfit175bestc14}.  \\
\indent Since imposing the twist does not seem to introduce a novel character to the TWR measurement, by extension this leads us to conclude that the regularized TWR calculation is insensitive to minor radial variation in the PA (i.e. $\delta_{PA}$=$\pm$5$\degree$ over roughly 4 kpc), if real.  However, if the disk PA is assertably closer to 175$\degree$ than 170$\degree$, this seems to reinforce the impression that the bright spiral structure may be best described by a single constant pattern speed.  \\
\indent Though compelling, we emphasize that this exercise should not be interpreted as confirmation or denial of radial variation in the PA, nor as providing an unequivocal measure for the pattern speed of the bright spiral structure.  Critically, imposing the twist tends to remove the most noticeable asymmetries in the velocity field, particularly within 60'', very much in the manner described previously for PA=175$\degree$. If true signatures of pattern speeds have been obscured or eliminated at this PA, this may prevent the measurement of a distinct pattern speed inside $r$$\sim$2.0 kpc.  
\subsection{\label{sec:discussion}Relation to $m$=3 structure}
\indent In principle, TWR measurements at either PA=170$\degree$ or 175$\degree$ may reflect signatures of patterns other than those of the bright two-armed structure alone.  The Fourier power spectrum of the surface brightness reveals rich structure in the disk of M51, much of it coexisting over roughly 2 kpc in the inner disk.  In this section we review the particular possibility that weak $m$=3 structure identified by \citet{rix} out to $\sim$2.5 kpc contributes with a unique pattern speed to out TWR solutions.\\
\indent Only if all structures in the same radial zone have identical pattern speeds will the TWR calculation accurately reflect this sole speed; given a surface brightness distribution which reflects two coincident contributions (say, from $m$=2 and $m$=3 structures) with unique time dependence (i.e. different pattern speeds), the TWR calculation is currently unequipped to constrain either one or the other. Though a generalization can be made under the assumption that both pattern speeds are constant, it is beyond the scope of this work to develop either the TW or TWR calculation appropriate to the situation.  \\
\indent For M51, it may be possible in the future to directly relate the velocity asymmetry from arm to arm of the bright $m$=2 spiral structure identified by \citet{henry} to the presence of the $m$=3 mode, and to its pattern speed in particular. (\citealt{henry} have already successfully demonstrated that the presence of the $m$=3 mode induces a systematic offset in the azimuthal positions of the two main arms).  This should allow us to establish the expected combination of speeds in the TWR calculation; if the implied $m$=3 pattern speed is unlike the measure $\Omega_{p,1}$ found with PA=170$\degree$, for example, this speed is presumably unshared by the $m$=2 structure.  
\section{\label{sec:conclusion}Conclusion}
In this paper we present regularized TWR solutions for the pattern speed of the bright spiral structure in the inner disk of M51, derived with velocity and intensity information from the ISM-dominant molecular component traced by high-resolution CO observations.  These solutions are arrived at by isolating the inner disk from errors which evidently originate with both the quality of sampling/detection and the pattern speed-modeling in bins covering the outer, material pattern.  So although our procedure prevents us from constraining the outermost pattern speed, calculating the outer bins without regularization in principle improves the accuracy with which the solution for the inner disk can realize the true pattern speed.  \\
\indent Our primary result with this implementation is the measurement inside 4 kpc of two pattern speeds, both significantly higher, and together fitting the data better, than the constant global measure of \citet{zrm04} at the nominal PA=170$\degree$.  A third, lower pattern speed, extending beyond 4 kpc out to (at least) 5.3 kpc is also detected, nearer the speed expected for the material pattern.  Significantly, the transitions between the measured pattern speeds coincide with resonances; the two pattern speeds inside 4 kpc both end at corotation within the uncertainties.  Since it is in no way imposed by the method, this dynamically reasonable scenario tends to give us confidence as to the physical plausibility of the pattern speeds returned by the analysis.  \\
\indent Of course, given that a pattern speed interior to $r$$\simeq$2 kpc is only weakly detected (if at all) at PA=175$\degree$, the accuracy of the description provided by two pattern speeds may depend on whether PA=170$\degree$ or PA=175$\degree$ is more accurate, an uncertainty raised by \citet{shetty}, for instance.  If the disk is best described with PA=175$\degree$, we find evidence that a single constant pattern speed inside 4 kpc best characterizes the bright spiral structure.  Furthermore, as contemplated in $\S$ \ref{sec:varyPA}, a radially varying PA which decreases from 180$\degree$ (and reaches 170$\degree$ near 3 kpc), perhaps suggested by the results of \citet{shetty}, also favors a single measured pattern speed interior to 4 kpc.  \\
\indent Again, however, while the analysis presented here cannot resolve the question as to which PA is more appropriate, we find meaningful, independent evidence in favor of the pattern speeds measured at 170$\degree$, in particular.  For example, consistent with expectations of leading structure at the inner most radii (as in the observations of \citealt{scoville}), the higher speed inside $r$$\simeq$2 kpc lacks an ILR.  In addition, attendant to our finding that both speeds interior to 4 kpc terminate at corotation, the transition between the two roughly coincides with an inferred location of resonance overlap wherein the inner's corotation resonance and the outer's inner 4:1 resonance align.  The radial domain of the pattern speed measured at PA=175$\degree$, in contrast, is not as clearly associated with resonance radii.  Since the bright spiral structure does not appear along the minor axis near $\sim$2 kpc, the corotation resonance at this location implied by the solution at 170$\degree$ is unfortunately unconfirmable through inspection of radial streaming velocities under the density wave interpretation.  \\
\indent We also find remarkable agreement between the characteristics of the two speeds inside 4 kpc at PA=170$\degree$ and other evidence in the inner disk consistent with multiple pattern speeds.  The transition parameterized in our best-fit solution for the inner disk coincides with significant variation in the two-armed spiral pitch angle.  Since a change in the pitch angle is expected to be accompanied by a change in streaming motions, both parameters are presumably attendant to the signatures (streaming or otherwise) of the patterns.  \\
\indent Although the pattern speed interior to 2 kpc in the solution at 170$\degree$ (or 175$\degree$) may reflect a unique contribution from the $m$=3 mode observed by \citet{rix}, as described in $\S$ \ref{sec:discussion}, the measurements in Figure \ref{fig-globalcut9a} (or Figure \ref{fig-twfit175bestc14}) presumably directly relate to the patterns present in the disk and so (depending on the PA) should provide a fair description of the dynamics therein.  As such, it may be possible that ensuing observations and studies better discriminate between the two seemingly disparate radial dependencies implied for the PAs considered here.  \\
\indent Even at present our TWR solutions yield interpretations with which to observationally address the question of spiral longevity.  That the solutions at both 170$\degree$ and 175$\degree$ feature constant pattern speeds would imply that our solutions are indicative of long-lasting spiral structure.  Interestingly, at the innermost radii both qualitatively resemble the model of \citet{salob} where, characteristic of the isolated evolution of the disk, the dominant $m$=2 component has a constant pattern speed $\sim$50 $\kmskpc$ out to $\sim$1.2-1.8 kpc.  As for the region between $\sim$1.8 kpc and 4.6 kpc in those models where interaction with the companion introduces a succession of transient structures, our solutions at both 170$\degree$ and 175$\degree$ otherwise describe at most two steady patterns in distinct radial zones.  \\
\indent In the immediate future, observations with higher resolution and sensitivity should afford TWR calculations with finer radial bins, thereby allowing for the parameterization of more distinct radial zones, if present.  This will either confirm our solutions for $\Omega_p(r)$ or perhaps demonstrate that solutions describe a succession of many discrete patterns (similar to the transient structures in the \citet{salob} models), or simply a winding, material pattern.   Again, though, our multiple-speed and other, quadratic solutions in general more closely follow $\Omega$ throughout the disk than $\Omega-\kappa/2$ characteristic of $m$=2 structure in the models of \citet{salob}.  \\
\indent Despite the lingering ambiguity in the PA, these TWR solutions present a new picture of the bright spiral structure of M51, one that should prompt tests of long-lived density wave theories in other nearby grand-design spirals.  At the very least, this study marks a successful starting point for continued tests of the relation between multiple spiral pattern speeds in a single disk; investigations into the number and radial domains of pattern speeds and spiral winding in nearby spirals will be the subject of upcoming work.  \\

S. E. M. acknowledges support from a NASA-funded New Mexico Space Grant Consortium Graduate Research Fellowship.  We thank Heikki Salo for providing insight into some of the intricacies of the simulations in \citet{salob}.



\begin{thebibliography}{}
\bibitem[Aalto et al.(1999)]{aalto}Aalto, S., Huttemeister, S., Scoville, N. Z. \& Thaddeus, P. 1999, ApJ, 522, 165
\bibitem[Bresolin et al.(2004)]{bresolin} Bresolin, F., Garnett, D. R., \& Kennicutt, R. C. 2004, ApJ, 615, 228
\bibitem[Debattista(2003)]{debPA}Debattista, V. P. 2003, MNRAS, 342, 1194
\bibitem[Debattista et al.(2006)]{debetal}Debattista, V. P., Mayer, L., Carollo, C. M., Moore, B., Wadsley, J., Quinn, T. 2006, ApJ, 645, 209
\bibitem[Elmegreen et al.(1989)]{elmegreen1}Elmegreen, B. G., Elmegreen, D. M. \& Seiden, P. E. 1989, ApJ, 343, 602
\bibitem[Faber \& Gallagher (1979)]{fabgall}Faber, S. M., \& Gallagher, J. S. 1979, ARAA, 17, 135
\bibitem[Garcia-Burillo et al.(1993)]{garcia}Garcia-Burillo, S., Combes, F., \& Gerin, M. 1993, A\&A, 274, 148
\bibitem[Henry et al.(2003)]{henry}Henry, A. L., Quillen, A. C. \& Gutermuth, R. 2003, AJ, 126, 2831
\bibitem[Howard \& Byrd(1990)]{howardByrd}Howard, S. \& Byrd, G. G. 1990, AJ, 99, 1798
\bibitem[Kennicutt(1989)]{kenn89}Kennicutt, R. 1989, ApJ, 344, 685
\bibitem[Kennicutt et al.(2007)]{kennicutt}Kennicutt, R., Calzetti, D., Walter, F., Helou, G., Hollenbach, D. J., Armus, L., Bendo, G., Dale, D. A., Draine, B. T., Engelbracht, C. W., Gordon, K. D., Prescott, M. K. M., Regan, M. W., Thornley, M. D., Bot, C., Brinks, E., de Blok, E., de Mello, D., Meyer, M., Moustakas, J., Murphy, E. J., Sheth, K., \& Smtih, J. D. T. 2007, ApJ, 671, 333
\bibitem[Lin(1970)]{lin}Lin, C. C. 1970, in IAU Symp. 38, The Spiral Structure of Our Galaxy, ed. W. Becker \& G. I. Contopoulos (Dordrecht: Reidel), 377
\bibitem[Lin \& Shu(1964)]{linshu}Lin, C. C. \& Shu, F. H. 1964, ApJ, 140, 646
\bibitem[Masset \& Tagger(1997)]{masset}Masset, F. \& Tagger, M. 1997, A\&A, 322, 442

\bibitem[Meidt et al.(2008)]{meidt}Meidt, S. E., Rand, R. J., Merrifield, M. R., Debattista, V. P. \& Shen, J. 2008, ApJ, 676, 899
\bibitem[Merrifield, Rand \& Meidt(2006)]{mrm}Merrifield, M. R., Rand, R. J., \& Meidt, S. E. 2006, MNRAS, 366, 17
\bibitem[Miller(1970)]{mill}Miller, K. 1970, SIAM Journal on Mathematical Analysis, Vol 1, 52
\bibitem[Patrikeev et al.(2006)]{patrikeev}Patrikeev, I., Fletcher, A., Stepanov, R. Beck, R., Berkhuijsen, E. M., Frick, P. \& Horellou, C. 2006, A\&A, 458, 441
\bibitem[Rand(1993)]{rand93}Rand, R. J. 1993, ApJ, 410, 68
\bibitem[Rand \& Wallin(2004)]{rw04}Rand, R. J. \& Wallin, J. F. 2004, ApJ, 614, 142
\bibitem[Rautiainen \& Salo(1999)]{rs}Rautiainen, P. \& Salo, H. 1999, A\&A, 348, 737
\bibitem[Rix \& Rieke(1993)]{rix}Rix, H. W. \& Rieke, M. J. 1993, ApJ, 418, 123
\bibitem[Salo \& Laurikainen(2000a)]{salo}Salo, H. \& Laurikainen, E. 2000, MNRAS, 319, 377
\bibitem[Salo \& Laurikainen(2000b)]{salob}Salo, H. \& Laurikainen, E. 2000, MNRAS, 319, 393
\bibitem[Scoville et al.(2001)]{scoville}Scoville, N. Z., Polletta, M., Ewald, S., Stolovy, S. R., Thompson, R. \& Rieke, M. 2001, AJ, 122, 3017
\bibitem[Shetty et al.(2007)]{shetty}Shetty, R., Vogel, S. N., Ostriker, E. C. \& Teuben, P. J. 2007, ApJ, 665,1138
\bibitem[Sofue et al.(1999)]{sofue}Sofue, Y., Tutui, Y., Honma, M., Tomita, A., Takamiya, T., Koda, J., \& Takeda, Y. 1999, ApJ, 523, 136
\bibitem[Sygnet et al.(1988)]{syg}Sygnet, J. F., Tagger, M., Athanassoula, E. \& Pellat, R. 1988, MNRAS, 232, 733
\bibitem[Tagger et al.(1987)]{tagger}Tagger, M., Sygnet, J. F., Athanassoula, E. \& Pellat, R. 1987, ApJL, 318, 43
\bibitem[Tikhonov \& Arsenin(1977)]{tich}Tikhonov, A. N. \& Arsenin, V. Y. 1977, Solutions of Ill-posed Problems (New York: Wiley)
\bibitem[Toomre(1981)]{toomre}Toomre, A. 1981, in The Structure and Evolution of Normal Galaxies, eds. S. M. Fall \& D. Lynden-Bell (London: Cambridge Univ. Press), 111
\bibitem[Toomre \& Toomre(1972)]{toomresq}Toomre, A. \& Toomre J. 1972, ApJ, 178, 623
\bibitem[Tully(1974)]{tully}Tully, R. B. 1974, ApJS, 27, 449
\bibitem[Tremaine \& Weinberg(1984)]{tw84}Tremaine, S. \& Weinberg, M. D. 1984, ApJ 282, L5
\bibitem[Vogel et al.(1993)]{vog93}Vogel, S. N., Rand, R. J., Gruendel, R. A. \& Teuben, P. J. 1993, PASP, 105, 60
\bibitem[Zaritsky, Rix \& Rieke(1993)]{zrr}Zaritsky, D., Rix, H.-W. \& Rieke, M. 1993, Nat, 364, 313
\bibitem[Zimmer, Rand \& McGraw(2004)]{zrm04}Zimmer, P., Rand, R. J. \& McGraw, J. T. 2004, ApJ, 607, 285

\end{thebibliography}
\end{document}